\documentclass[preprint,12pt]{elsarticle}

\usepackage{graphicx}

\usepackage{amssymb, amsthm}

\biboptions{compress}

\journal{\hspace{-3.5cm} \raisebox{-1mm}{\begin{tikzpicture} \draw [fill = white, white] (0, 0) rectangle (3.5, 0.4); \end{tikzpicture}}}

\usepackage{amsmath}
\usepackage{url}
\usepackage{color}
\usepackage{tensor}
\usepackage{mathrsfs}
\usepackage{nicefrac}
\usepackage{tikz-cd}
\usepackage{tikz}
\usetikzlibrary{decorations.markings}
\usetikzlibrary{matrix,arrows}
\usepackage{array}
\usepackage{ulem}
\usepackage{placeins}

\usepackage{pdflscape}

\usepackage[latin1]{inputenc}      
\usepackage[T1]{fontenc}    
\usepackage{ae,aecompl}                          
\usepackage{dsfont}
\usepackage{appendix}
\usepackage{amsfonts,amsmath,latexsym,bbm}
\usepackage{amssymb}
\usepackage{accents}
\usepackage{graphics}
\usepackage{graphicx}
\usepackage{booktabs,longtable}
\usepackage{bm}
\usepackage{dcolumn}
\usepackage{enumitem}
\usepackage{upgreek}
\usepackage{tensor}

\usepackage{tikz}
\usepackage{tikz-cd}
\usepackage{tikz-3dplot}
\usetikzlibrary{decorations.markings}
\usetikzlibrary{matrix,arrows}
\usetikzlibrary{calc}
\usetikzlibrary{shapes}
\usetikzlibrary{positioning}
\usetikzlibrary{fit}

\usepackage{pgfplots}
\pgfplotsset{width=7cm,compat=1.5.1}

\usetikzlibrary{decorations.pathreplacing}
\makeatletter
\def\underbrace#1{%
  \@ifnextchar_{\tikz@@underbrace{#1}}{\tikz@@underbrace{#1}_{}}}
\def\tikz@@underbrace#1_#2{%
  \tikz[baseline=(a.base)] {\node[inner sep=2] (a) {\(#1\)};
  \draw[thick,line cap=round,decorate,decoration={brace,amplitude=4pt}]
    (a.south east) -- node[pos=0.5,below,inner sep=7pt] {\(\scriptstyle #2\)} (a.south west);}}
\makeatother

\usepackage[linktocpage=true]{hyperref}

\makeatletter
\renewcommand*{\eqref}[1]{%
  \hyperref[{#1}]{\textup{\tagform@{\ref*{#1}}}}%
}
\makeatother

\setlist{font=\normalfont\itshape} 

\renewcommand{\vec}[1]{\boldsymbol{#1}}
\newcommand{\tsr}[1]{\overset\leftrightarrow{#1}}
\newcommand{\ext}{_{\rm ext}}
\newcommand{\tot}{_{\rm tot}}
\newcommand{\ind}{_{\rm ind}}

\newcommand{\h}{\hspace{1pt}}
\newcommand{\hh}{\hspace{0.5pt}}
\newcommand{\mh}{\hspace{-1pt}}
\newcommand{\mhh}{\hspace{-0.5pt}}
\renewcommand{\j}{\mathrm i}
\newcommand{\de}{\mathrm d}

\newcommand{\lar}[1]{\textnormal{\mbox{\large $#1$}}}
\renewcommand{\i}{\mathrm i}

\definecolor{oldgray}{gray}{0.4}

\newcommand{\T}{_{\mathrm T}}
\renewcommand{\L}{_{\mathrm L}}

\DeclareMathAlphabet{\mathbbmsl}{U}{bbm}{m}{sl}

\numberwithin{equation}{section}

\begin{document}

\begin{frontmatter}



\title{Wavevector-{\itshape dependent} optical properties from wavevector-{\itshape independent} proper conductivity tensor}


\author[freiberg]{R.~Starke}
\ead{starke.ronald@googlemail.com}

\author[freiberg]{R.~Wirnata\corref{cor1}}
\ead{Rene.Wirnata@physik.tu-freiberg.de}

\author[aachen]{G.\,A.\,H.~Schober}
\ead{schober@physik.rwth-aachen.de}

\author[freiberg]{N.~Bulut\corref{cor1}}
\ead{bulutnbht@gmail.com}

\author[freiberg]{J.~Kortus}
\ead{Jens.Kortus@physik.tu-freiberg.de}

\cortext[cor1]{Corresponding author.}

\address[freiberg]{Institute for Theoretical Physics, TU Bergakademie Freiberg, Leipziger Stra\ss e 23, \\ 09599 Freiberg, Germany \vspace{0.1cm}}
\address[aachen]{Institute for Theoretical Solid State Physics, RWTH Aachen University, Otto-Blumenthal-Stra\ss e, 52074 Aachen \vspace{0.1cm}}

\begin{abstract}
We discuss the calculation of the refractive index by means of the ab initio {\itshape scalar dielectric function} and point out its inherent limitations. 
To overcome these, we start from the recently proposed fundamental, microscopic wave equation in materials
in terms of the frequency- {\itshape and} wave\-{}vector-dependent {\itshape dielectric tensor,} and investigate under which
conditions the standard treatment can be justified. Thereby, we address the question of neglecting the wavelength dependence of microscopic response functions. 
Furthermore, we analyze in how far the fundamental, microscopic wave equation is equivalent 
to the standard wave equation used in {\itshape theoretical optics}. In particular, we clarify the relation of the ``effective'' dielectric
tensor used there to the microscopic dielectric tensor defined in {\itshape ab initio physics.}
\end{abstract}
\begin{keyword}
electronic structure \sep electrodynamics in media \sep refractive index \sep optical activity \sep birefringence \\[6pt]
{\it Cite as:~} ResearchGate pub.~{\bfseries 319207955}; arXiv:1708.06330; conventionally published as: Eur.~Phys.~J.~B, 93 3 (2020) 54


\end{keyword}

\end{frontmatter}



\newpage
\tableofcontents

\bigskip
\section{Introduction} \label{sct:Intro}

More often than not, optical material properties are accessed in {\itshape ab initio materials physics} via the standard relation 
between the {\itshape macroscopic} dielectric function $\varepsilon(\omega)$ and the refractive index $n(\omega)$ given by\footnote{Here, as in Ref.~\cite{EDWegner}, we suppress the subscript ``$\rm r$'' of the {\itshape relative} dielectric function and simply write $\varepsilon \equiv \varepsilon_{\rm r}$\h.} (see Refs.~\cite[Eq.~(8.33)]{Strinati}, \cite[Eq.~(2.17)]{Hanke}, \cite[Eqs.~(18.26)]{Bechstedt}, \cite[Eq.~(6.11)]{Cardona},\cite[p.~534]{Ashcroft} or \cite[Eq.~(2.203)]{SchafWegener}) \smallskip
\begin{equation}
 n^2(\omega) = \varepsilon(\omega) \,. \smallskip \label{eq_standard}
\end{equation}
The macroscopic dielectric function, for its part, is defined as the limit $\vec k \to \vec 0$ 
of the microscopic (frequency- {\itshape and} wavevector-dependent) dielectric function 
(see Refs.~\cite[Eqs.~(8.25)]{Strinati}, \cite[Eq.~(2.18)]{Hanke} or \cite[Eq.~(2.207)]{SchafWegener}):
\begin{equation}
 \varepsilon(\omega) = \lim_{|\vec k|\rightarrow 0} \varepsilon(\vec k,\omega) \,. \label{eq_direction}
\end{equation}
Importantly, the dielectric function used in these equations is typically calculated from the density response function, and hence it
corresponds to the {\itshape longitudinal} part of the dielectric tensor
(see Refs.~\cite[Eq.~(18.23)]{Bechstedt}, \cite[\S\,2.6.4]{SchafWegener}, \cite[Vol.~1, \S\,4]{NozieresPinesBook}).
Although the standard treatment delivers sensible results for a huge variety of materials 
(see Refs.~\cite{Berger,Lee,Laszlo,Kawai,Makhnev,Rusinov15,Akrap} for recent examples), it ultimately proves insufficient.
To make this clear, let us summarize the main conceptual problems of this standard treatment based on the {\itshape wavevector-independent} dielectric {\itshape function:}
\begin{enumerate}
\item As a matter of principle, optical  properties correspond to {\itshape transverse} (although not necessarily {\itshape purely} transverse)
electromagnetic waves in a material, and hence they should  not be deduced from a purely {\itshape longitudinal} response function (at least not in a na\"ive way).
\item The standard relation for the refractive index, Eq.~\eqref{eq_standard}, is only valid in the limit $\vec k \to \vec 0$, whereas light
waves definitely have a non-vanishing wavevector, i.e., $\vec k\neq \vec 0$.
\item In particular, for anisotropic media, the refractive index should  at least depend {\itshape on the direction} of the wavevector $\vec k$.
Correspondingly, the limit \eqref{eq_direction} actually depends on the direction (cf.~Ref.~\cite[Eq.~(18.22)]{Bechstedt} or discussion below)
\item Moreover, for {\itshape birefringent} or {\itshape optically active} materials, there are {\itshape pola\-{}rization-dependent} refractive indices.
Accordingly, the joint determination of polarization vectors {\itshape and} refractive indices corresponding to a given wavevector obviously requires a tensorial equation.
By contrast, Eq.~\eqref{eq_standard} yields at best one refractive index, whose polarization vector cannot---even in principle---be defined for vanishing wavevectors.
\end{enumerate}
We particularly stress that the shortcomings of the standard relation between the dielectric function and the refractive index are not discovered by the authors of this article but are in fact well-known and acknowledged in the theoretical literature, where the validity of Eq.~\eqref{eq_standard} is typically restricted to cubic crystals (see e.g.~Refs.~\cite[p.~32]{Strinati}, \cite[p.~294]{Hanke}, \cite[p.~432]{Bechstedt}). This does, however, not stop people from applying it to other systems as well (see Refs.~\cite{dongho_nguimdo_density_2015,gracia_characterization_2009,loper_complex_2015,saha_electronic_2000,friedrich_optical_2017,yamada_time-dependent_2018}).

Befittingly, a recent, meritorious article by D.~Sangalli {\itshape et al.} \cite{Myrta} has drawn attention to the fact that optical properties
can also be calculated from the {\itshape wavevector-dependent} current response {\itshape tensor}. On the other hand,
the above considerations make it clear that in general, the calculation of optical material properties requires the wave- {\it and} the polarization vector to be taken into account,
and this in turn requires a treatment based on the wavevector dependent current response tensor (or equivalently, the dielectric tensor \cite{EDWegner,Forcella,Fiorino}).
In fact, from the dielectric tensor---which naturally contains much more information than the {\itshape scalar} dielectric function---the density response function can 
be reconstructed by means of the Universal Response Relations \cite{ED1,ED2,Refr},
while the converse is not true (see Refs.~\cite{Strinati,Giuliani,Altland,Melrose1Book}), i.e.~the dieletric tensor cannot be reconstructed from the dielectric function, although this is sometimes claimed. 

However, in the limit $\vec k \to \vec 0$, such relations between different response functions cannot be evaluated due to their singular behavior (see, for example, Eq.~\eqref{eq_EpsStandardRel} below, which relates the dielectric function  to the density response function). Correspondingly, the authors of Ref.~\cite{Myrta} have based their treatment on a full {\itshape ab initio} calculation
of the wavevector-dependent response functions. While this is certainly the right approach in the most general case, 
the downside is that such wavevector-dependent calculations are in general extremely demanding,
in particular if they are supposed to yield dispersion relations.

In this article, we resume this problem on a fundamental level to show that it is possible---at least in principle---to obtain wavevector-dependent optical properties without calculating wavevector-dependent response functions numerically, provided that the treatment is based on the assumption that the {\itshape proper conductivity tensor} is
given in the optical limit. Fortunately, this is exactly what is common practice
in numerical studies using state of the art electronic structure codes like
e.g.~Elk task 121 \cite{elk}. Thus, our considerations are predicated on the presupposition that the conductivity tensor has already been calculated using the appropriate ab initio methods, which for their part are not addressed in this article. Put differently, we are not concerned in this article with the improvement of the ab initio calculations, but consider the wavevector (or at least wavelength) independent conductivity tensor as an input typically provided for by some ab initio electronic structure code. We furthermore have to stress, in this context,
that the condition of wavelength independence cannot simultaneously apply to all other response functions as well (see the discussion below).

This article is organized as follows: In \S\,\ref{Sct:Resp}, we assemble some general relations between wavevector-dependent response functions.
In \S\,\ref{Sct:waveEqs}, we discuss the fundamental, microscopic wave equation in materials in terms of the proper conductivity tensor, 
and we investigate the conditions under which the treatment of optical material properties by means of a scalar dielectric function can be justified. In particular, we define the {\itshape effective dielectric tensor} in terms of the proper conductivity tensor, and we show that by rewriting the fundamental wave equation in terms of this effective dielectric tensor, the resulting equation agrees formally with the standard wave equation used in theoretical optics.
The subsequent \S\,\ref{Sct:OpticalProp} is dedicated to an in-depth analysis of this wave equation and its solutions under the assumption of a wavelength-independent proper conductivity tensor. Finally, in \S\,\ref{appendix} we reproduce some well-known facts from theoretical optics within our general formalism by assuming special forms for the effective dielectric tensor.

\section{Response functions and wavevector dependence}\label{Sct:Resp}

Our investigations are based on the {\it functional approach to electrodynamics of materials} \cite{EDWegner,ED1,ED2,Refr,EDOhm,EffWW,EDLor,EDWave,EDFullGF,EDFresnel}, which is a paradigmatically new approach motivated by the common practice in modern ab initio materials science (see extensive discussion in Refs.~\cite{ED1,ED2,Refr}). In particular, the functional approach omits macroscopic averaging procedures and is hence intrinsically microscopic. Correspondingly, it replaces the traditional distinction of ``free'' and ``bound'' source terms in the Maxwell equations with the modern distinction of {\it external} and {\it internal} (or {\it induced}\h) fields as used in first principles theory. Consequently, the traditional ``constitutive'' equations assumed to characterize a material can be identified with {\it response relations} mediating between external and induced quantities, which in turn paves the way for the application of the Kubo formalism (see e.g.~Refs.~\cite[Sct.~3]{Giuliani}, \cite[Sct.~6]{Bruus}) so conspicuously absent from the traditional textbook literature on electrodynamics in media. On the practical side, the functional approach to electrodynamics of materials is then primarily concerned with the deduction of interrelations between different electromagnetic response functions from purely electromagnetic considerations. Entre autres, this leads to a whole number of {\it Universal Response Relations}, which are entirely unknown in the traditional approach, while the ab initio community is at least partially familiar with them.

In fact, for purely classical electrodynamic reasons it turns out that the {\itshape current response tensor} $\chi$ already contains the complete information about all linear, electromagnetic response properties. E.g., it is related to the conductivity tensor $\sigma$ by means of a Universal Response Relation \cite[\S\,6]{ED1} reading (see Refs.~\cite[Eqs.~(2.177) and (2.198)]{SchafWegener}, \cite[Eq.~(3.185)]{Giuliani}, \cite{EDOhm}, and for a gauge-independent derivation see Ref.~\cite[\S\,3.2.3]{ED2}):
\begin{equation}
 \tsr{\chi}(\vec x, \vec x';t-t') = -\partial_t \h \tsr{\sigma}(\vec x, \vec x'; t-t') \,. \label{eq_uni1}
\end{equation}
The current response tensor for its part is the spatial part of the {\itshape fundamental response tensor} \cite[\S\,7.4]{Altland},
\begin{equation} \label{chimn}
\chi\indices{^\mu_\nu}(x,x') = \frac{\delta j^\mu\ind(x)}{\delta A^\nu\ext(x')}\,,
\end{equation}
where $j^\mu = (c \rho, \vec j)^{\rm T}$ denotes the (induced) electromagnetic four-current density, and $A^\nu = (\varphi/c, \vec A)^{\rm T}$ the (external) four-potential. As is evident from Eq.~\eqref{eq_uni1}, microscopic response functions constitute in general entire tensor valued integral kernels (as opposed to sheer numbers). In particular, they are typically {\it non-local} and {\it inhomogeneous}. For the latter fact, there exist two fundamentally different reasons on the theoretical level. Firstly, real probes are finitely extended in space thereby possessing a specific geometry. Consequently, their response functions cannot be spatially homogeneous already for that reason. In theoretical materials science, however, this aspect can be discarded because here we are concerned with purely material dependent (and hence geometry independent) response functions, which would---at least conceptually---correspond to probes filling out all of space homogeneously. Concretely, this means that the response functions have to be calculated in the {\it thermodynamic limit}. Even so, putting aside such heuristic model systems as the homogeneous electron gas, the microscopic response function is not homogeneous due to the atomic structure of the material. Rather, in the case of {\it crystalline} materials it turns out that the Fourier transform of a typical electromagnetic response function, say the density response function, is of the general form
\begin{equation}
 \upchi = \upchi_{\vec G\vec G'}(\vec k,\omega) \,,
\end{equation}
where $\vec G$ and $\vec G'$ are reciprocal lattice vectors while $\vec k$ is a vector in the first Brillouin zone. Applying this to fields whose Fourier transforms are supported in the first Brillouin zone anyway, we obtain the particularly simple response relation
\begin{equation}
 \rho_{\rm ind}(\vec k,\omega) = \upchi_{\vec 0\vec 0}(\vec k,\omega)\varphi_{\rm ext}(\vec k,\omega) \,,
\end{equation}
whose simplistic multiplicative structure translates into
\begin{equation}
 \rho_{\rm ind}(\vec x,t) = c\int\de t'\int\de\vec x'\,\upchi_{\vec 0\vec 0}(\vec x - \vec x';t-t')\varphi_{\rm ext}(\vec x',t')
\end{equation}
in the space-time domain. Hence, in this case the response function appears to be strictly homogeneous (i.e.,~it depends exclusively on the {\it difference} of the spatial variables $\vec x$ and $\vec x'$). Correspondingly, we call the transition $\vec G,\vec G'\rightarrow \vec 0$ in a response relation the {\it homogeneous limit}. As, on the other hand, the modulus of a wavevector in the first Brillouin zone roughly fulfills the approximate inequality $|\vec k|\lessapprox \pi/a $ with a typical lattice constant $a$, the said homogeneous limit corresponds to the condition that the involved wavelengths are larger than at least twice the lattice constant, $2a\lessapprox\lambda$. This is of course a version of the famous Nyquist-Shannon sampling theorem, where the external field corresponds to the signal while the crystal plays the r\^{o}le of the sampler. 

In practice, i.e.~in optical experiments, the involved wavelengths are even much larger than typical lattice constants and hence it is highly intuitive that at such wavelengths the material---even if crystalline on the microscopic level---appears to be homogeneous on the macroscopic level. The {\it homogeneous limit} can therefore be identified with the traditional {\it transition to macroscopic fields}, where it has to be borne in mind though that in our case the fields do not have to be averaged but are instead simply assumed to lie in the, say, optical range (which they certainly do in the case of those experiments which are our concern here). Instead of the fields, it is therefore now the response function which is to be subjected to such a macroscopic transition, which, however, is not implemented anymore by some complicated averaging procedure (as it was in the traditional approach), but by the evaluation at small wavevectors in the sense of $\vec G,\vec G'\rightarrow\vec 0$. 

Interestingly, in the said homogeneous limit, the fundamental response tensor is now of the following general form (see e.g.  Refs.~\cite{Strinati,ED1,Altland,Melrose1Book}):
\begin{equation}\label{generalform1}
\chi^\mu_{~\nu}(\vec k,\omega)=
\left( \!
\begin{array}{rr} -\lar{\frac{c^2}{\omega^2}} \, \vec k^{\rm T} \, \tsr{\chi}(\vec k,\omega)\h\hh \vec k & \lar{\frac{c}{\omega}} \, \vec k^{\rm T} \, \tsr{\chi}(\vec k,\omega)\, \\
[12pt] -\lar{{\frac{c}{\omega}}} \, \tsr{\chi}(\vec k,\omega)\h\hh \vec k & \, \tsr{\chi}(\vec k,\omega)\, 
\end{array} \right).
\end{equation}
In particular, the {\itshape density response function} $\upchi$
can be calculated from the current response tensor as follows \cite[Eq.~(3.175)]{Giuliani}:
\begin{equation}
 \upchi(\vec k,\omega) \h := \h \frac{\delta \rho_{\rm ind}(\vec k, \omega)}{\delta \varphi_{\rm ext}(\vec k, \omega)} \h = \h \frac 1 {c^2} \h \chi\indices{^0_0}(\vec k, \omega) \h = \h -\frac{\vec k^{\rm T} \h \tsr{\chi}(\vec k,\omega) \h \vec k}{\omega^2} \,. \label{eq_uni2}
\end{equation}
We remark that the response relations \eqref{eq_uni1}--\eqref{eq_uni2} hold
analogously for the respective {\itshape proper} response functions, which relate the induced quantities to the total (i.e., external plus induced) quantities (see Ref.~\cite[\S\,2.3]{Refr}).
Furthermore, the {\itshape proper} density response function is related to the dielectric function by the equation \cite[Eq.~(2.172)]{SchafWegener}
\begin{equation}
 \varepsilon(\vec k,\omega) = 1 - v(\vec k) \h\hh \widetilde\upchi(\vec k,\omega) \,, \label{eq_EpsStandardRel}
\end{equation}
where $v(\vec k) = 1 / (\varepsilon_0 |\vec k|^2)$ denotes the Coulomb interaction kernel in Fourier space. 
We stress again that the dielectric function in this equation actually coincides 
with the {\itshape longitudinal} part of the dielectric tensor (see Ref.~\cite[\S\,2.6.4]{SchafWegener}) given by
\begin{equation}
 \varepsilon(\vec k,\omega) \equiv \varepsilon_{\rm L}(\vec k,\omega)  
 = \frac{\vec k^{\rm T} \h\hh \tsr{\varepsilon}(\vec k,\omega) \h\hh \vec k}{|\vec k|^2} \,. \label{eq_genDefDielFct}
\end{equation}
With this, Eq.~\eqref{eq_EpsStandardRel} can be proven directly by applying the functional chain rule (see Refs.~\cite{ED1} and \cite[\S\,5.1]{EDWave}). More directly, it can be shown as follows: The relation of the dielectric tensor to the proper current response tensor reads \cite[Eq.~(3.41)]{EDWave}
\begin{equation}
 \tsr{\varepsilon}(\vec k,\omega) = \tsr 1 - \tsr{D}_0(\vec k,\omega) \tsr{\widetilde\chi}(\vec k,\omega) \,,
\end{equation}
where
\begin{equation}
 \tsr{D}_0(\vec k,\omega) = {\mathbbmsl D}_0(\vec k,\omega) \left(\tsr 1 - \frac{c^2|\vec k|^2}{\omega^2} \tsr P\L \right) \,,
\end{equation}
and
\begin{equation}
 {\mathbbmsl D}_0(\vec k,\omega) = \frac{c^2\mu_0}{-\omega^2 + c^2|\vec k|^2} \,.
\end{equation}
We then get (skipping some straightforward algebra)
\begin{align}
 \varepsilon\L(\vec k,\omega)  &= \frac{\vec k^{\rm T}\,\tsr\varepsilon(\vec k,\omega)\vec k}{|\vec k|^2} \\
                &= 1 - {\mathbbmsl D}_0(\vec k,\omega) \frac{\vec k^{\rm T}}{|\vec k|^2} \left(\tsr 1 - \frac{c^2|\vec k|^2}{\omega^2}\tsr P\L \right)\tsr{\widetilde\chi}(\vec k,\omega)\vec k \\
                &= 1 - \frac{c^2\mu_0}{|\vec k|^2} \bigg(- \frac{\vec k^{\rm T}\tsr{\widetilde\chi}(\vec k,\omega)\vec k}{\omega^2}\bigg) \\
                &= 1 - \frac{\widetilde\upchi(\vec k,\omega)}{\varepsilon_0|\vec k|^2} \,,
\end{align}
which shows the assertion, Eq.~\eqref{eq_EpsStandardRel}. The importance of this proof lies in the fact that the usual arguments in favor of Eq.~\eqref{eq_EpsStandardRel} are either based on the isotropic limit or operate exclusively with longitudinal fields. Correspondingly, it is a priori unclear whether the general definition of the dielectric function given in Eq.~\eqref{eq_genDefDielFct} lends itself to the standard relation \eqref{eq_EpsStandardRel}. 

Importantly, as stressed already, the microscopic current response tensor---or equivalently, the microscopic conductivity tensor---already 
contains the complete information about all linear electromagnetic response properties (this insight can be traced back at least to Ref.~\cite{KeldyshKirzhnitz}; 
the fact as such has also been stressed recently in Ref.~\cite{Forcella}; 
for a systematic derivation of all linear electromagnetic response functions in terms of the conductivity tensor see also Ref.~\cite[\S\,5--6]{ED1}). 
In particular, it is possible to reconstruct from the microscopic conductivity tensor the response with respect to strictly longitudinal perturbations. 
However, one has to emphasize as well that the corresponding response relations are formulated in terms of response functions at finite wavevectors $|\vec k|>0$ (see Ref.~\cite{Myrta}).
Hence, these relations can in general not be evaluated na\"{i}vely in the limit $\vec k \to \vec 0$, but require precise knowledge about the response functions 
in the vicinity of the origin and the direction of its approach.

Finally, we remark again that the response relations presented above hold in this form only for homogeneous materials, 
while they become more involved in the general (i.e., inhomogeneous) case \cite{ED1,EDLor}. Moreover, as opposed to the (longitudinal) dielectric function such quantities as the {\it transverse} dielectric function are actually meaningful only in the isotropic limit (see Refs.~\cite[App.~D.1]{EffWW}, \cite[\S\,5.1]{EDWave}, \cite[\S\,2.1]{EDFullGF}).

\section{Linear wave equations in materials}\label{Sct:waveEqs}
\subsection{Fundamental, microscopic wave equation} \label{Subsct:genwave}

On a fundamental level, the most general, linear electromagnetic wave equation in materials---which requires
only spatial homogeneity---reads as follows (see discussions in Refs.~\cite{EDWegner,Refr,EDWave,Dolgov,EDFresnel}):
\begin{equation}
 \tsr\varepsilon(\vec k,\omega) \h \vec E(\vec k,\omega) = 0 \,. \label{eq_general}
\end{equation}
Here, $\tsr\varepsilon(\vec k, \omega)$ denotes the {\itshape ab initio} dielectric tensor, which is {\it defined} by the linear approximation (see Refs.~\cite[Eq.~(2.140)]{SchafWegener},
\cite[Eq.~(5.198) and (5.203)]{Kantorovich}, and \cite[Eq.~(E.10)]{Martin}),
\begin{equation}
 \vec E\ext (\vec k,\omega) \overset{\rm def}{=} \tsr\varepsilon(\vec k,\omega) \h\vec E\tot(\vec k,\omega)   \,, \label{eq_fundDef} \smallskip
\end{equation}
where we will set from now on $\vec E \equiv \vec E\tot$. Although the {\itshape ab initio} derivation of the fundamental wave equation \eqref{eq_general} is somewhat complicated \cite[\S\,4.1]{Refr}, its ultimate meaning is just that the external field does not penetrate the material or, put differently, 
that the electromagnetic waves in the material correspond to the latter's {\itshape proper oscillations}.

In the case of an isotropic medium, the fundamental wave equation decouples into
\begin{align}
 \varepsilon_{\rm L}(\vec k, \omega) \h \vec E_{\rm L}(\vec k, \omega) & = 0 \,, \label{eq_long} \\[5pt]
 \varepsilon_{\rm T}(\vec k, \omega) \h \vec E_{\rm T}(\vec k, \omega) & = 0 \,, \label{eq_trans}
\end{align}
which are equations formulated in terms of the {\itshape longitudinal} and {\itshape transverse} dielectric function.
The resulting conditions for the existence of nontrivial solutions,
\begin{align}
 \varepsilon_{\rm L}(\vec k,\omega_{\vec k\rm L}) &= 0 \,, \label{eq_long_disp} \\[5pt]
 \varepsilon_{\rm T}(\vec k,\omega_{\vec k\rm T}) &= 0 \,, \label{eq_trans_disp}
\end{align}
determine the respective {\itshape dispersion relations}, $\omega_{\vec k{\rm L}}$ and $\omega_{\vec k{\rm T}}$\hh,
of the electromagnetic proper oscillations \cite[Eq.~(2.34)]{Dolgov}. In the longitudinal case described by Eq.~\eqref{eq_long_disp}, 
the resulting waves are conventionally called {\itshape plasmons} (see e.g.~Refs.~\cite[Eq.~(5.49)]{Giuliani}, \cite[Eq.~(14.78)]{Bruus}, and \cite[Eq.~(4.92)]{MartinRothen}).
The corresponding {\itshape transverse proper oscillations} determined by 
Eq.~\eqref{eq_trans_disp}---whose existence can already be deduced {\itshape per analogiam} from Eq.~\eqref{eq_long_disp}---obviously 
describe the propagation of light in the medium. In the most general (i.e., not necessarily isotropic) case, however, both equations combine into the unified
wave equation \eqref{eq_general}. It remains to discuss in how far this fundamental, microscopic wave equation translates into
the standard wave equation used in theoretical optics.

\subsection{Wave equation used in theoretical optics}

For the above purpose, we now reformulate the general wave equation \eqref{eq_general} in terms of the microscopic {\itshape proper} conductivity tensor \cite[\S\,2.3]{Refr}.
As a matter of principle \cite[\S\,2.5]{Refr}, this quantity is related to the dielectric tensor via \cite[Eqs.~(2.24)--(2.25)]{Dolgov}
\begin{equation}\label{cond_rel_2}
 \tsr\varepsilon(\vec k, \omega) = \tsr 1 - \tsr{\mathbbmsl E}(\vec k, \omega) \, \frac{\tsr{\widetilde \sigma}(\vec k, \omega)}{\j\omega \h \varepsilon_0} \,,
\end{equation}
where the {\itshape electric solution generator} \cite[\S\,2.4]{Refr} is given in terms of 
the well-known {\itshape longitudinal} and {\itshape transverse projection operators} \cite[Eqs.~(2.1) and (2.2)]{EDWave} by the concise formula
\begin{equation}\label{eq_sol_gen}
 \tsr {\mathbbmsl E}(\vec k, \omega) = \tsr P_{\mathrm L} (\vec k) + \frac{\omega^2}{\omega^2 - c^2 |\vec k|^2} \, \tsr P_{\mathrm T}(\vec k) \,.
\end{equation}
Eq.~\eqref{cond_rel_2} holds independently of the isotropic limit. It is, in fact, quite generally valid, i.e.~regardless of the approximation used to calculate the conductivity tensor. 
With these preparations, the fundamental wave equation \eqref{eq_general} can be rewritten as
\begin{equation}
 \tsr {\mathbbmsl E}{}^{-1}(\vec k, \omega) \h \vec E(\vec k,\omega) = \frac{1}{\j\omega \h \varepsilon_0} \h \tsr{\widetilde \sigma}(\vec k, \omega) \h \vec E(\vec k,\omega) \,,
\end{equation}
and further using Eq.~\eqref{eq_sol_gen}, we can recast it into the form
\begin{equation}\label{eq_transWaveEq}
 \bigg({-\frac{\omega^2}{c^2}} \h \bigg( \hh \tsr 1- \frac{\tsr{\widetilde\sigma}(\vec k,\omega)}{\i\omega \h \varepsilon_0}\bigg) + |\vec k|^2 \h \tsr P\T(\vec k) \bigg) \vec E(\vec k,\omega) = 0 \,. 
\end{equation}
Finally, by applying the vector identity
\begin{equation}
 \vec k \times \big(\vec k \times \vec E(\vec k,\omega)\big) = -|\vec k|^2 \h \tsr P\T(\vec k) \h \vec E (\vec k,\omega) \,,
\end{equation}
and by defining the {\itshape effective dielectric tensor} as
\begin{equation}\label{eq_defEffDiTens}
 \tsr\varepsilon_{\rm eff}(\vec k,\omega) \h \overset{\rm def}{=} \h \tsr 1 - \frac{\tsr{\widetilde \sigma}(\vec k, \omega)}{\j\omega \h \varepsilon_0}  \,,
\end{equation}
we can bring Eq.~\eqref{eq_transWaveEq} into the equivalent form
\begin{equation}
 {-\frac{\omega^2}{c^2}}\h\tsr\varepsilon_{\rm eff}(\vec k,\omega) \h \vec E(\vec k,\omega) = \vec k \times \big (\vec k \times \vec E(\vec k,\omega) \big ) \,. \label{eq_transWaveEqMod}
\end{equation}
This equation {\itshape formally agrees} with the standard wave equation used in {\itshape theoretical optics} and solid state physics 
(see e.g.~Refs.~\cite[Eq.~(1)]{Draxl16}, \cite[Eq.~(4.11)]{Roemer}, \cite[Eq.~(2.2.9)]{Agranovich}, and \cite[Eq.~(41.2)]{Platzmann}).
We emphasize, however, that the {\itshape effective} dielectric tensor appearing in Eq.~\eqref{eq_transWaveEqMod} is actually not identical, 
but only approximately equal (in the limit $|\vec k|\rightarrow \vec 0$) to the {\itshape real} (i.e., {\itshape ab initio}) dielectric tensor defined in Eq.~\eqref{eq_fundDef} (compare Eqs.~\eqref{cond_rel_2} and \eqref{eq_defEffDiTens}).
Put differently, we here observe an astonishing ``error cancellation'': the {\itshape phenomenological} wave equation \eqref{eq_transWaveEqMod} combined with the {\itshape approximate} relation between the dielectric tensor and the conductivity
tensor, Eq.~\eqref{eq_defEffDiTens}, is {\itshape precisely equivalent} to the {\itshape exact} wave equation \eqref{eq_general} in the form of Eq.~\eqref{eq_transWaveEq}.
This explains, in particular, why the {\itshape ab initio} treatment of the refractive index, if based on Eq.~\eqref{eq_transWaveEq},
will be {\itshape entirely correct} even though it does not explicitly start from the fundamental wave equation \eqref{eq_general} (compare also Ref.~\cite{Dolgov}, Eqs.~(2.29)--(2.30), (2.33)--(2.34), and comments therein, where this insight is formulated---apparently for the first time---for the case of an isotropic dielectric function).

Next, we consider again the isotropic limit: from Eqs.~\eqref{cond_rel_2}--\eqref{eq_sol_gen}, it follows that Eq.~\eqref{eq_defEffDiTens} holds exactly for the longitudinal dielectric function \cite[Eq.~(E.11)]{Martin}, i.e.,
\begin{equation}
 \varepsilon_{\rm L}(\vec k, \omega) \h = \h 1 - \frac{\widetilde \sigma_{\rm L}(\vec k, \omega)}{\j\omega \h \varepsilon_0} \h =\h  \varepsilon_{\rm eff, \hh L}(\vec k, \omega) \,. \label{long_relation}
\end{equation}
On the other hand, for the transverse parts we find the relation\footnote{The identities \eqref{long_relation} and \eqref{trans_relation} can be compared to Ref.~\cite[Eqs.~(52)  and (53)]{EDWegner}; in particular, the phenomenological model defined in Ref.~\cite{EDWegner} has the property that $\varepsilon_{\rm eff, \hh L}(\vec k, \omega) = \varepsilon_{\rm eff, \hh T}(\vec k, \omega) = \varepsilon_{\rm eff}$
with a {\itshape constant, scalar} effective permittivity $\varepsilon_{\rm eff}$\h.}
\begin{equation}
 \varepsilon_{\rm T}(\vec k, \omega)\h  =\h  1 - \frac{\omega^2}{\omega^2 - c^2 |\vec k|^2} \, \frac{1}{\j \omega \h \varepsilon_0} \, \widetilde \sigma_{\rm T}(\vec k, \omega)\h =\h \frac{\varepsilon_{\rm eff, \hh T}(\vec k, \omega) \, \omega^2 - c^2 |\vec k|^2}{\omega^2 - c^2 |\vec k|^2} \,. \label{trans_relation}
\end{equation}
Furthermore, Eq.~\eqref{eq_transWaveEqMod} decouples in the isotropic limit into two separate wave equations for the longitudinal and transverse electric field components:
\begin{align}
 \varepsilon_{\rm eff, \hh L}(\vec k, \omega) \h \vec E_{\rm L}(\vec k, \omega) & = 0 \,, \\[5pt]
 \bigg( {-\frac{\omega^2}{c^2} \h \varepsilon_{\rm eff, \hh T}(\vec k, \omega)} + |\vec k|^2 \bigg) \hh \vec E_{\rm T}(\vec k, \omega) & = 0 \,. \label{eq_compare}
\end{align}
Together with Eqs.~\eqref{long_relation}--\eqref{trans_relation}, these equations are in fact equivalent to Eqs.~\eqref{eq_long}--\eqref{eq_trans} from the previous subsection. This explains, 
in particular, {\itshape why in {\itshape ab initio} physics one usually works with two entirely different wave equations,}
both formulated in terms of the dielectric function: the phenomenological wave equation \eqref{eq_transWaveEqMod}, 
which is usually used for optical, i.e., transverse oscillations only, 
and the plasmon equation \eqref{eq_long}, which is used  for longitudinal oscillations.
In actual fact, however, by using the {\itshape exact} relation \eqref{cond_rel_2} between the dielectric tensor and the conductivity tensor, 
both wave equations turn out to be of the same type; and in case that longitudinal
and transverse oscillations do not decouple, they have to be combined into the fundamental wave equation \eqref{eq_general}.

Finally, we also comment on the wave equation corresponding to the standard equation \eqref{eq_standard} for the refractive index, which is \cite[Eq.~(2.203)]{SchafWegener}
\begin{equation}
 \bigg({-\frac{\omega^2}{c^2}} \h \varepsilon_{\rm L}(\omega) + |\vec k|^2 \bigg) \h \vec E_{\rm T}(\vec k,\omega) = 0 \,. \label{eq_approximate}
\end{equation}
This does clearly not coincide with Eq.~\eqref{eq_compare}. Astonishingly, Eq.~\eqref{eq_approximate} constitutes instead a wave equation for a {\itshape transverse} electric field, which is, however, formulated in terms of the {\itshape longitudinal} dielectric function. 
Nevertheless, this equation can also be justified from the fundamental wave
equation in the isotropic case, i.e., from Eqs.~\eqref{eq_long}--\eqref{eq_trans}: At optical wavelengths (i.e., for small wavevectors), it is sometimes plausible to assume that the longitudinal and transverse conductivities coincide in the sense of
\begin{align}\label{eq_longAppr}
\widetilde\sigma\L(\vec k,\omega) = \widetilde\sigma\T(\vec k,\omega) \,.
\end{align}
With this relation, one shows directly from Eq.~\eqref{cond_rel_2} that the longitudinal and transverse dielectric functions are related as follows \cite[\S\,4.4]{Refr}:
\begin{equation}\label{eq_RelLongTans}
  \varepsilon_{\rm T}(\vec k,\omega) = \frac{-\omega^2 \h \varepsilon_{\rm L}(\vec k,\omega) + c^2|\vec k|^2}{-\omega^2 + c^2|\vec k|^2} \,,
\end{equation}
i.e.~provided the approximation \eqref{eq_longAppr} holds, the transverse dielectric function can be calculated from its longitudinal counterpart by this formula. Together with the fundamental wave equation \eqref{eq_trans}, this implies
\begin{equation}
 \bigg( {-\frac{\omega^2}{c^2}}\h \varepsilon_{\rm L}(\vec k,\omega) + |\vec k|^2 \bigg) \h \vec E_{\rm T}(\vec k,\omega) = 
 \bigg({-\frac{\omega^2}{c^2}} + |\vec k|^2 \bigg) \h \varepsilon_{\rm T}(\vec k,\omega) \h \vec E_{\rm T}(\vec k,\omega) = 0 \,.
\end{equation}
Finally, by neglecting the $\vec k$ dependence of $\varepsilon_{\rm L}$ we arrive at  the assertion, Eq.\linebreak \eqref{eq_approximate}. On the other hand, since the condition \eqref{eq_longAppr} 
will not always be true, and since the $\vec k$ dependence of $\varepsilon_{\rm L}$ cannot always be neglected, our results corroborates~again our initial statement (see \S\,\ref{sct:Intro}) that in the most general case,
the deduction of optical material properties should  not be based on the standard relation \eqref{eq_standard}.

\section{Optical properties from conductivity tensor}\label{Sct:OpticalProp}
\subsection{Fundamental wave equation in optical limit}

The above considerations have shown that as a matter of principle, optical properties have to be deduced from the general wave equation encompassing the 
transverse subspace, i.e., from Eq.~\eqref{eq_general}, or from its standard form given by Eq.~\eqref{eq_transWaveEqMod}, which can also be written as
\begin{equation} \label{implicit}
  \tsr\varepsilon_{\rm eff}(\vec k, \omega) \h \vec E(\vec k, \omega) = \frac{c^2 |\vec k|^2}{\omega^2} \, \tsr P_{\rm T}(\vec k)\h \vec E(\vec k, \omega) \,.
\end{equation}
Unfortunately, this equation requires (via the relation \eqref{eq_defEffDiTens}) knowledge of the full, i.e., 
frequency- and wavevector-dependent conductivity tensor, which is computationally very demanding. Moreover, 
even if the full conductivity tensor or the corresponding effective dielectric tensor is known, the dispersion relation $\omega = \omega_{\vec k}$ will have to be deduced from the {\itshape implicit} equation \eqref{implicit}, where the frequency appears not only explicitly on the right-hand side but also implicitly as an argument 
of the effective dielectric tensor. Consequently, the refractive index, which is defined by (see Refs.~\cite[Eq.~(1.2.5)]{Agranovich} or \cite[Eq.~(11.12)]{Melrose}) \smallskip
\begin{equation}
n_{\vec k} = \frac{c \hh |\vec k|}{\omega_{\vec k}} \,, \label{eq_defRefrInd} \smallskip \vspace{2pt}
\end{equation}
is also only given by an implicit equation (see Ref.~\cite[\S\,4.2]{Refr}).

The decisive point is now that instead of interpreting Eq.~\eqref{implicit} as an implicit equation determining the frequency (or the refractive index) as a function of the wavevector, 
one may instead also fix the frequency and the {\itshape direction} of the wavevector, $\hat{\vec k} := \vec k / |\vec k|$, and regard Eq.~\eqref{implicit} 
as an implicit equation determining the modulus of the wavevector $|\vec k|$. Correspondingly, one can also consider the refractive index as a function of 
the frequency and the {\it direction} of the wavevector, $n = n(\hat{\vec k}, \omega)$. The modulus of the wavevector is then given in terms of this refractive index by
\begin{equation}
 |\vec k| = \frac{\omega}{c} \, n(\hat{\vec k}, \omega) \,. \label{eq_ModWaveVec}
\end{equation}
Even in this case, though, the refractive index is given only implicitly by Eq.\ \eqref{implicit}, because 
the modulus $|\vec k|$ does not only appear explicitly on the right-hand side of this equation
but also implicitly as an argument of the effective dielectric tensor (see also discussion in Sct.~\ref{Sct:Wavelength} below). 

A pragmatic way to overcome these difficulties lies in the following assumption: 
although the relations between different response functions are in general {\itshape wavevector dependent} (see \S\,\ref{Sct:Resp}), 
it is not contradictory to assume that {\itshape one particular} 
response function is actually {\itshape wavelength independent} (at least approximately), 
whereby it has to be stressed that this assumption {\itshape cannot} be upheld {\itshape simultaneously}
for all response functions at once (see again \S\,\ref{Sct:Resp}, or the most general Universal Response Relations in Ref.~\cite[\S\,6]{ED1}).
In particular, the wavelength independence of the conductivity tensor implies that
the density response function as well as the dielectric tensor are definitely wavelength dependent (see Eqs.~\eqref{eq_uni2} and \eqref{cond_rel_2}).

Concretely, using the relation between the wavelength and the modulus of the wavevector, 
\begin{equation}
|\vec k| = \frac{2\pi}{\lambda} \,,
\end{equation}
we see that our assumption of wavelength independence concretely means that the dependence on the modulus can be dropped in the above formulary. In fact, the standard approach suggests that the assumption of wavelength independence actually applies to the {\itshape proper conductivity tensor}, or equivalently (see Eq.~\eqref{eq_uni1}) to the proper current response tensor. The justification for this claim is that the calculation of a genuine dielectric tensor (as opposed to a shear dielectric function as in Eq.~\eqref{eq_EpsStandardRel}) is, as by Eq.~\eqref{eq_defEffDiTens}, in practice based on the determination of the conductivity tensor in the first place, which for its part typically comes as a completely wavevector independent computer output. Put differently, in practical calculations one may even assume that the proper conductivity tensor is wavevector independent altogether, i.e.~$\widetilde\sigma=\widetilde\sigma(\omega)$, though conceptually this is actually overly restrictive, for which reason we do not insist on this point here. Correspondingly, for now (see Sct.~\ref{Sct:Wavelength} below for the more general case) we assume that at optical wavelengths, the {\itshape proper} conductivity~tensor is wavelength independent in the sense that
\begin{align} \label{cond_cond}
 \tsr{\widetilde\sigma}(\vec k, \omega) \equiv \tsr{\widetilde\sigma}(\hat{\vec k},|\vec k|,\omega) = \tsr{\widetilde \sigma}(\hat{\vec k},\omega) \,.
\end{align}
In the following, we will investigate the general wave equation under this additional assumption. We remark in passing, though, that the proper conductivity tensor can of course trivially be identified with its limit $|\vec k|\rightarrow 0$ (or equivalently $\lambda\rightarrow\infty$)---which defines the so-called {\itshape optical limit}---if it is actually wavelength independent anyway.

First, Eq.~\eqref{cond_cond} implies that the effective dielectric tensor defined by Eq.\ \eqref{eq_defEffDiTens} also depends exclusively on the frequency.
This, in turn, greatly simplifies the solution of the wave equation \eqref{implicit}, which now becomes an {\itshape explicit} equation determining 
the modulus of the wavevector $|\vec k|$ as a function of $\hat{\vec k}$ and $\omega$. In fact,
since the transverse projection operator for its part depends only on the direction (and not on the modulus) of the wavevector,
\begin{equation} \label{def_trans_proj}
 \tsr P_{\rm T}(\vec k) = \tsr 1 - \frac{\vec k \hh \vec k^{\rm T}}{|\vec k|^2} = \tsr P_{\rm T}(\hat{\vec k})\,,
\end{equation}
the {\itshape frequency- and direction-dependent refractive indices} are now determined by an {\itshape explicit} equation,
whose directional dependence comes into play only via the transverse projection operator. To show this, we make the following ansatz for the electric field in Fourier space:
\begin{equation}
\begin{aligned}\label{eq_ansatz}
 \vec E(\vec k,\omega) & \propto \vec e(\hat{\vec k},\omega) \, \delta \mh\mhh \left(|\vec k|-\frac{\omega}{c}\, n(\hat{\vec k}, \omega)\right) \\[5pt]
 & \quad \, + \vec e^*(-\hat{\vec k},-\omega) \, \delta \mh\mhh \left(|\vec k|+\frac{\omega}{c} \, n(-\hat{\vec k},-\omega)\right) ,
\end{aligned}
\end{equation}
where $\vec e(\hat{\vec k},\omega)$ is the so-called {\itshape polarization vector}, which we assume to be normalized. 
Note that the polarization vector may, in principle, also depend on the frequency (though {\itshape in praxi} it is usually frequency independent). The ansatz \eqref{eq_ansatz} fulfills the constraint condition
\begin{equation}
    \vec E(\vec k,\omega) = \vec E^*(-\vec k, -\omega) \,,
\end{equation}
which guarantees that the electric field is real-valued in the space-time domain. Now, by putting Eq.~\eqref{eq_ansatz} into Eq.~\eqref{implicit}, 
we see that our ansatz solves the general wave equation provided that the polarization 
vector fulfills the {\itshape central equation for the joint determination of refractive indices and polarization vectors:}
\begin{equation} \label{eq_central}
  \tsr\varepsilon_{\rm eff}(\hat{\vec k},\omega) \, \vec e(\hat{\vec k},\omega) = n^2(\hat{\vec k},\omega) \, \tsr P_{\rm T}(\hat{\vec k})\, \vec e(\hat{\vec k},\omega) \,,
\end{equation}
as well as the analogous condition
\begin{equation} \label{eq_central_2}
  \tsr\varepsilon_{\rm eff}(\hat{\vec k},\omega) \, \vec e^*(-\hat{\vec k},-\omega) = n^2(-\hat{\vec k},-\omega) \, \tsr P_{\rm T}(\hat{\vec k})\,\vec e^*(-\hat{\vec k},-\omega) \,.
\end{equation}
Furthermore, the reality condition on the effective dielectric tensor,
\begin{equation}
 \tsr\varepsilon_{\rm eff}(\hat{\vec k},\omega) = (\tsr\varepsilon_{\rm eff})^*(-\hat{\vec k},-\omega) \,,
\end{equation}
and the trivial fact that
\begin{equation}
 \tsr P_{\rm T}(\hat{\vec k}) = (\tsr P_{\rm T})^*(-\hat{\vec k}) \,,
\end{equation}
together imply that the condition \eqref{eq_central_2} is actually equivalent to Eq.~\eqref{eq_central} and can therefore be discarded. On the other hand, by
combining Eqs.~\eqref{eq_central} and \eqref{eq_central_2} one sees directly that
\begin{align}
 \vec e(\hat{\vec k},\omega) &= \vec e^*(-\hat{\vec k},-\omega) \,, \\[3pt]
 n^2(\hat{\vec k}, \omega) & = n^2(-\hat{\vec k}, -\omega) \,,
\end{align}
and since Eq.~\eqref{eq_ModWaveVec} should always yield a positive result for the modulus of the wavevector \cite{Veselago}, we may even set 
\begin{equation}
n(\hat{\vec k},\omega) = -n(-\hat{\vec k},-\omega) \,.
\end{equation}
In summary, Eq.~\eqref{eq_central} is the central equation for the joint determination 
of the frequency- and direction-dependent refractive indices and the corresponding polarization vectors. The logic of this central equation is this: the  frequency and the direction of the wavevector are independent
variables which, in principle, can be prescribed arbitrarily; these being given, Eq.\ \eqref{eq_central} determines the refractive index and via Eq.~\eqref{eq_ModWaveVec} the modulus of the wavevector, as well as the possible polarization vectors of the proper oscillations with the given frequency and direction.

Finally, we remark that Eq.~\eqref{implicit} is {\itshape formally analogous} to the equation $\vec D = \varepsilon_0 \h n^2 \vec E_{\rm T}$\h, 
which is known from theoretical optics (see Refs.~\cite[\S\,15.2, Eqs.~(2) and (4)]{BornWolf} and \cite[Eq.~(6.21)]{Lipson}), provided
one plugs in the standard relation 
\begin{equation}
\vec D \overset{\rm def}{=} \varepsilon_0\tsr\varepsilon_{\rm eff}\vec E \label{eq_defDispl}
\end{equation}
for the electric ``displacement'' field and uses Eq.~\eqref{eq_defRefrInd}. In fact, in our context Eq.~\eqref{eq_defDispl} has to be interpreted as the definition of the classical field $\vec D$. Appropriately, the displacement field then also fulfills the standard equation
\begin{equation}
 \nabla\cdot\vec D = \rho_{\rm ext} \equiv \rho_{\h\text{``free''}} \,,
\end{equation}
as follows from
\begin{align}
 \nabla\cdot\vec D(\omega) &= \varepsilon_0\nabla\cdot\bigg(\vec E(\omega) - \frac{\tsr{\widetilde \sigma}(\omega)}{\j\omega \h \varepsilon_0}\vec E(\omega)\bigg) \\
&=\varepsilon_0\nabla\cdot\left(\vec E(\omega) - \frac{\vec j_{\rm ind}(\omega)}{\j\omega \h \varepsilon_0}\right) \\ 
&=\rho(\omega) - \frac{\nabla\cdot\vec j_{\rm ind}(\omega)}{\j\omega} \\
&=\rho_{\rm ext}(\omega) \,,
\end{align}
where we have used 
\begin{equation}
\rho\equiv\rho_{\rm tot}=\rho_{\rm ext}+\rho_{\rm ind} = \rho_{\h\text{``free''}} + \rho_{\h\text{``bound''}} 
\end{equation}
and the continuity equation in Fourier space,
\begin{equation}
 -\i\omega\rho(\vec x,\omega) + \nabla\cdot\vec j(\vec x,\omega) = 0 \,.
\end{equation}
By contrast, the transverse part of the displacement field, $\nabla\times\vec D_{\rm T} =\nabla\times\vec D$, depends on the conductivity tensor under consideration. We note, however, that the thus defined displacement field would be a purely auxiliary quantity (though not a particularly helpful one) without the slightest intrinsic meaning. In particular, it would be of the most secondary character as it could only be calculated ex post, i.e.~once the conductivity tensor is given, and can hence not serve any practical purpose. Worse, it does not even have a value on the conceptual level, e.g.~as a defining equation for the effective dielectric tensor. Rather to the contrary, the latter must be used to define the displacement field in the first place.

The present
article therefore works within the paradigm of the Functional Approach to electrodynamics in media (see Refs.~\cite{ED1,ED2} for extensive discussions), 
which is an ab initio theory exclusively based on the microscopic Maxwell equations. 
Instead of introducing ``macroscopic'' fields, it simply distinguishes between external and induced
fields as it is common practice in ab initio materials physics (see e.g.~the modern textbooks \cite{SchafWegener,Giuliani,Altland,Bruus}).
Note, however, that one cannot---in general---apply the standard methods of theoretical optics to \eqref{implicit},
because the effective dielectric tensor---stemming from a {\itshape retarded} response function calculated by the Kubo formula---is typically {\itshape not hermitean} 
(as it is assumed in theoretical optics) and hence not necessarily diagonalizable. 
In the following, we will study in detail the solutions of Eq.~\eqref{eq_central}
{\itshape without} any particular assumption about the effective dielectric tensor.

\subsection{Radiation modes and generalized plasmons}

Although Eq.~\eqref{eq_central} is not just an eigenvalue problem for the effective dielectric tensor, it allows for 
a somewhat analogous mathematical treatment: nontrivial solutions exist if and only if the condition (see Refs.~\cite[Eq.~(4.12)]{Roemer}, \cite[Eq.~(77.9)]{Landau} and \cite[Eq.~(2.14)]{Zvezdin})
\begin{equation} \label{det_cond}
 \det \mh \left( \h \tsr \varepsilon_{\rm eff}(\hat{\vec k},\omega) - n^2 (\hat{\vec k}, \omega) \h \tsr P_{\rm T}(\hat{\vec k}) \right) = 0
\end{equation}
is fulfilled. As stressed above, for given direction $\hat{\vec k}$ and frequency $\omega$, this equation determines the 
refractive indices $n_{\lambda}(\hat{\vec k}, \omega)$, which we label by an index $\lambda \in \mathbb N$ (later, we will see that $\lambda \in \{1, 2\}$).
These solutions can be studied most conveniently by choosing for each direction~$\hat{\vec k}$ an orthonormal basis in $\mathbb R^3$, i.e., 
three (real) vectors $\vec e_{\vec k1}$, $\vec e_{\vec k2}$, $\vec e_{\vec k3}$ with the property that $\vec e_{\vec k i} \cdot \vec e_{\vec kj} = \delta_{ij}$\h. 
We further assume that $\vec e_{\vec k1}$ and $\vec e_{\vec k2}$ 
are perpendicular to $\vec k$, i.e., they lie in the transverse subspace, while $\vec e_{\vec k3} = \hat{\vec k}$ is in the (one-dimensional) longitudinal subspace. We call this an {\it isotropic basis}. Eq.~\eqref{det_cond} then takes the following~form:
\begin{equation} \label{det_mat}
 \det \mh \left( \begin{array}{lll} \varepsilon_{11} - n^2 & \varepsilon_{12} & \varepsilon_{13} \\[3pt] \varepsilon_{21} & \varepsilon_{22} - n^2 & \varepsilon_{23} \\[3pt] \varepsilon_{31} & \varepsilon_{32} & \varepsilon_{33} \end{array} \right) = 0 \,,
 \end{equation}
where we have defined the {\it optical tensor}, i.e.~the matrix corresponding to the effective dielectric tensor evaluated in the isotropic basis
\begin{align}
  \varepsilon_{ij}(\hat{\vec k}, \omega) \h \overset{\rm def}{=} \h \vec e_{\vec k i}^{\rm T} \, \tsr \varepsilon_{\rm eff}(\hat{\vec k},\omega) \, \vec e_{\vec k j} \,.
\end{align}
We emphasize particularly that these components of the {\itshape effective} dielectric tensor refer to a basis in $\vec k$-space (and hence not to
a fixed basis in real space). In particular, this implies that the matrix appearing in Eq.~\eqref{det_mat}
does in fact depend on the direction of $\vec k$, {\it even if the original effective dielectric tensor is purely frequency dependent} (as it often is). {\it This explains in particular why it is possible (and in fact imperative) to obtain wavevector dependent optical properties from wavevector independent computer outputs}.

Given a refractive index  $n_\lambda$ which solves Eq.~\eqref{det_mat}, one further obtains the (normalized) vectors $\vec v_{\lambda} = (v^1_\lambda, v^2_\lambda, v^3_\lambda)^{\rm T}$ 
which fulfill the homogeneous equation \smallskip
\begin{equation}\label{eq_central5}
 \left( \begin{array}{lll} \varepsilon_{11} - n^2_\lambda & \varepsilon_{12} & \varepsilon_{13} \\[3pt] \varepsilon_{21} & \varepsilon_{22} - n^2_\lambda & \varepsilon_{23} \\[3pt] \varepsilon_{31} & \varepsilon_{32} & \varepsilon_{33} \end{array} \right) \left( \! \begin{array}{l} v^1_\lambda \\[3pt] v^2_\lambda \\[3pt] v^3_\lambda \end{array} \! \right) = 0 \,. \smallskip
\end{equation}
The corresponding (normalized) polarization vectors, which solve Eq.~\eqref{eq_central}, can then be written as
\begin{equation}
 \vec e_{\lambda}(\hat{\vec k}, \omega) = v_\lambda^1(\hat{\vec k}, \omega) \, \vec e_{\vec k1} + v_\lambda^2(\hat{\vec k}, \omega) \, \vec e_{\vec k2} + v_\lambda^3(\hat{\vec k}, \omega) \, \vec e_{\vec k3} \,,
\end{equation}
and hence they are, in general, {\itshape not} purely transverse (and possibly complex at that). 

In the remainder of this subsection, we will deduce some general properties of the solutions of Eqs.~\eqref{det_mat} and \eqref{eq_central5}. In the next subsection, we will then derive explicit expressions for the refractive indices in the most general case of an anisotropic material, and finally in \S\,\ref{appendix} we will investigate some special~cases. 

The first important observation concerning Eq.~\eqref{det_mat} is that it leads---as a consequence of the transverse projection operator appearing on the right-hand side of Eq.~\eqref{eq_central}---to a polynomial equation of {\itshape second} order in $n^2$, and hence there are (for each direction and frequency) at most {\itshape two} (possibly complex\footnote{Actually, our theoretical formalism is predicated on the assumption that the refractive index is real (see Eq.~\eqref{eq_ModWaveVec}). Nonetheless, one might accept complex refractive indices as an outcome of the final formulae if one does not care too much about how one has arrived at these in the first place. This, however, leaves open the question of what these complex refractive indices really mean, and this problem in turn is in no way specific for the stance proposed here. We therefore feel free to ignore it for the time being.}) refractive indices, which we denote by $n_{\lambda}(\hat{\vec k}, \omega)$ with $\lambda \in \{1, 2\}$. This is in stark contrast to an ordinary eigenvalue problem for a $(3 \times 3)$-matrix, which would in general have three solutions. 

However, the sole fact that there are at most two different refractive indices does not answer the question of how many
{\itshape radiation modes} exist in the medium with a given frequency and direction. Hereby, we define a radiation mode as a solution $(n^2(\hat{\vec k}, \omega), \h \vec e(\hat{\vec k}, \omega))$ of the central Eq.~\eqref{eq_central} with the following properties: (i) the refractive index is non-zero, $n^2(\hat{\vec k}, \omega) \not = 0$, and (ii) the polarization vector has a non-vanishing transverse part, $\vec e_{\rm T}(\hat{\vec k}, \omega) \not = 0$. 
The first condition is necessary because by Eq.~\eqref{eq_ModWaveVec}, a vanishing refractive index would imply that $|\vec k| = 0$. This means that the medium oscillates as a whole, a behavior which is also sometimes referred to as ``plasmon''. The second condition excludes the purely longitudinal proper oscillations of the medium, which are usually referred to as {\itshape plasmons} (see comments below). In other words, the question is now the 
following: for a given frequency and direction, how many linearly independent polarization vectors exist which solve Eq.~\eqref{eq_central} and which are not purely longitudinal? In analogy to the vacuum case, one might assume that there are at most two such modes for each direction and frequency (see \S\,\ref{app:vacuum}).
In order to prove this hypothesis, we take again recourse to theoretical optics (see e.g.~Refs.~\cite[\S\,4.2]{Roemer} and \cite[p.~300]{Bredov}), whereby we distinguish two cases depending on the determinant of the effective dielectric tensor.

\bigskip \noindent
{\itshape Case 1:} The effective dielectric tensor is invertible, hence $\mathrm{det}\,\tsr \varepsilon_{\rm eff} \not = 0$. In this case, acting on Eq.~\eqref{eq_central} first with the inverse effective dielectric tensor and then with the transverse projector, we obtain
\begin{equation}
 \left( \h\tsr P\T(\hat{\vec k}) \, (\tsr\varepsilon_{\rm eff})^{-1}(\hat{\vec k},\omega) \, \tsr P\T(\hat{\vec k})\right) \mh \vec e_{\rm T}(\hat{\vec k},\omega) = \frac{1}{n^2(\hat{\vec k}, \omega)} \, \vec e\T(\hat{\vec k},\omega) \,,
\end{equation}
where $\vec e_{\mathrm T}(\hat{\vec k}, \omega) = P\T(\hat{\vec k}) \h \vec e(\hat{\vec k}, \omega)$.
In the {\itshape transverse} subspace, this now {\itshape is} an eigenvalue problem, which shows that the (transverse parts of the) polarization vectors
can be characterized as eigenvectors of a suitably defined $(2\times2)$ matrix. The transverse part of the polarization vector being given,
we can then calculate the longitudinal part by the explicit formula
\begin{equation}
\vec e\L(\hat{\vec k},\omega) = n^2(\omega,\hat{\vec k}) \, (\tsr\varepsilon_{\rm eff})^{-1}(\hat{\vec k},\omega) \, \vec e\T(\hat{\vec k},\omega) -\vec e\T(\hat{\vec k},\omega) \,,
\end{equation}
which follows again from the central Eq.~\eqref{eq_central}.
Consequently, there are at most two linearly independent polarization vectors which possess a 
transverse part and thereby qualify as radiation modes.

\bigskip \noindent
{\itshape Case 2:} The effective dielectric tensor is not invertible, hence $\mathrm{det}\,\tsr \varepsilon_{\rm eff}  = 0$. In this case, one obvious solution of Eq.~\eqref{det_mat} is $n^2 = 0$, which does not qualify as a radiation mode. 
The other refractive index may be non-zero, and consequently, 
the possible polarization vectors are determined by the null space defined in Eq.~\eqref{eq_central5} with only one possible refractive index $n^2 \not = 0$. 
This null space could in principle even be three-dimensional; however, in this case the only non-vanishing components of the effective dielectric tensor would be $\varepsilon_{11} = \varepsilon_{22} = n^2$\h, and hence there would again be two transverse and one longitudinal oscillation.
Thus, even in the case of a singular effective dielectric tensor, there are at most two radiation modes.

It remains to discuss whether the kernel of the dielectric tensor has a physical meaning. For this purpose, we consider the condition
\begin{equation}\label{eq_plasmon}
 \tsr\varepsilon_{\rm eff}(\hat{\vec k},\omega) \, \vec e(\hat{\vec k},\omega) = 0 \,.
\end{equation}
If this equation is supposed to give rise to a proper oscillation of the medium, then the central equation \eqref{eq_central}
has to be fulfilled as well. A comparison shows that in this case, we either have $n^2(\hat{\vec k}, \omega)=0$ or $\vec e\T(\hat{\vec k}, \omega) = 0$.
The first possibility is not considered here. By contrast, the second possibility
states that the proper oscillation is purely longitudinal and hence corresponds 
to a so-called {\itshape plasmon}. Thus, Eq.~\eqref{eq_plasmon}
combined with the longitudinality condition constitutes the {\itshape generalized plasmon equation,} which generalizes the well-known condition \eqref{eq_long_disp} for isotropic media. In particular, since the effective dielectric tensor does not necessarily have a non-trivial kernel, this shows that plasmons do not necessarily have to exist in any material.

\subsection{General formulae for refractive indices}

In this subsection, we study the refractive indices in the most general case of an anisotropic material, 
for which the off-diagonal components of the (effective) dielectric tensor do not vanish. In this case, Eq.~\eqref{det_mat} leads to the following equation, which is quadratic in $n^2$\hh: 
\begin{equation} \label{poly}
 n^4 \, \varepsilon_{33} - n^2 \h \big( (\varepsilon_{11} + \varepsilon_{22}) \, \varepsilon_{33} - \varepsilon_{13} \, \varepsilon_{31} - \varepsilon_{23} \, \varepsilon_{32} \h \big) + \det \tsr \varepsilon_{\rm eff} = 0 \,.
\end{equation}
In theoretical optics, this is sometimes referred to as the
{\itshape Fresnel equation} (in honor of the legendary Augustin Jean Fresnel (1788--1827)) (see e.g.~Refs.~\cite[Eq.~(2.14)]{Zvezdin} or \cite[p.~300]{Bredov}). 
Note, however, that this equation must not be confused with the Fresnel equation{\itshape s} used for the intensity distribution for reflection at a material interface (see Ref.~\cite{EDFresnel}
for a recent discussion).

For studying the solutions of Eq.~\eqref{poly}, we distinguish again two cases depending on the value of $\varepsilon_{33}$\h. The latter parameter coincides, via the relation (cp.~Eq.~\eqref{eq_genDefDielFct})
\begin{equation}
 \varepsilon_{33}(\hat{\vec k}, \omega) \h \tsr P_{\rm L}(\hat{\vec k}) \h = \h \tsr P_{\rm L}(\hat{\vec k}) \h \tsr \varepsilon_{\rm eff}(\hat{\vec k},\omega) \h \tsr P_{\rm L}(\hat{\vec k}) \h \equiv \h (\tsr \varepsilon_{\rm eff})_{\rm LL}(\hat{\vec k}, \omega) \,,
\end{equation}
with the longitudinal projection (see Ref.~\cite[\S\,2.1]{EDWave}) of the effective dielectric tensor.

\bigskip \noindent
{\itshape Case a:} It may happen that at the given frequency and direction, we obtain $\varepsilon_{33}(\hat{\vec k}, \omega) = 0$. 
For such frequencies, Eq.~\eqref{poly} reduces to
\begin{equation}
 n^2 \h (\varepsilon_{13} \, \varepsilon_{31} + \varepsilon_{23} \, \varepsilon_{32} ) + \det \tsr \varepsilon_{\rm eff} = 0 \,,
\end{equation}
and this equation has {\itshape precisely one} solution $n^2$ given by
\begin{equation}
 n^2 = -\frac{\det \tsr \varepsilon_{\rm eff}}{\varepsilon_{13} \, \varepsilon_{31} + \varepsilon_{23} \, \varepsilon_{32}} \,.
\end{equation}
Here, we have assumed that the denominator does not vanish (which will generally be the case for anisotropic materials). 

\bigskip \noindent
{\itshape Case b:} $\varepsilon_{33}(\hat{\vec k}, \omega) \not = 0$. Now, Eq. \eqref{poly} has generally two (possibly complex) solutions given by
\begin{equation} \label{exact_sol}
\begin{aligned}
 (n^2)_{1/2} & = \frac 1 2 \left( \varepsilon_{11} + \varepsilon_{22} - \frac{\varepsilon_{13} \, \varepsilon_{31} + \varepsilon_{23} \, \varepsilon_{32}}{\varepsilon_{33}} \right) \\[5pt]
 & \quad \, \pm \frac 1 2 \,\h \sqrt{ \left(\varepsilon_{11} + \varepsilon_{22} - \frac{\varepsilon_{13} \, \varepsilon_{31} + \varepsilon_{23} \, \varepsilon_{32}}{\varepsilon_{33}} \right)^{\!\!2}  - \frac{4\det\tsr\varepsilon_{\rm eff}}{\varepsilon_{33}}} \,. 
\end{aligned}
\end{equation}
For frequencies $\omega = \omega(\hat{\vec k})$ which satisfy
\begin{equation}
 \det \tsr \varepsilon_{\rm eff}(\hat{\vec k}, \omega) = 0 \,,
\end{equation}
these solutions turn into
\begin{equation}
 n^2_1 = \varepsilon_{11} + \varepsilon_{22} - \frac{\varepsilon_{13} \, \varepsilon_{31} + \varepsilon_{23} \, \varepsilon_{32}}{\varepsilon_{33}}
 \end{equation}
 and $n^2_2 = 0$, where the latter refractive index can be discarded. (Thus, we recover Case 2 treated in the previous subsection.)

To summarize, Eqs.~\eqref{eq_central5} and \eqref{exact_sol} solve the problem of calculating the refractive indices
and polarization vectors in the most general case of a possibly non-diagonalizable effective dielectric tensor. We emphasize again that all the above formulae refer to the {\itshape effective} dielectric tensor defined by 
Eq.~\eqref{eq_defEffDiTens}. This can be calculated from the wave\-{}vector-{\itshape independent} proper conductivity tensor, but still allows one to deduce wavevector-{\itshape dependent} optical material properties from first principles.

\subsection{Wavelength dependent conductivity tensor}\label{Sct:Wavelength}

In the preceding subsections, we have assumed that the proper conductivity tensor is effectively wavelength independent at least in the optical range. Although this will certainly constitute an assumption which is fulfilled in almost all cases of practical relevance, we will shortly consider the possible failure of this assumption in this final subsection, if maybe only for conceptual reasons. In fact, if the proper conductivity tensor indeed perceptibly depends on the wavelength or---equivalently---on the modulus of the wavevector, so will the effective dielectric tensor, such that Eq.~\eqref{det_cond} has to be replaced by
\begin{equation} \label{det_condmod}
 \det \mh \left( \h \tsr \varepsilon_{\rm eff}(\hat{\vec k},|\vec k|,\omega) - n^2 (\hat{\vec k}, \omega) \h \tsr P_{\rm T}(\hat{\vec k}) \right) = 0 \,.
\end{equation}
At first glimpse, it now appears as if this equation was not sufficient for the determination of the optical properties anymore on the grounds  that it involves yet another variable, namely the said modulus of the wavevector. Fortunately, though, it still holds true that the modulus of the wavevector is not independent of its direction and the frequency. Instead, by means of the refractive index it can be expressed in terms of these via Eq.~\eqref{eq_ModWaveVec}. Plugging this back into the dielectric tensor would again turn Eq.~\eqref{det_condmod} into a closed, though implicit equation for the determination of the refractive indices,
\begin{equation} \label{det_condmodmod}
 \det \mh \left( \h \tsr \varepsilon_{\rm eff}\left(\hat{\vec k},\frac{\omega}{c}n(\hat{\vec k},\omega),\omega\right) - n^2 (\hat{\vec k}, \omega) \h \tsr P_{\rm T}(\hat{\vec k}) \right) = 0 \,.
\end{equation}
Superficially, it seems that this problem would be amenable to the methods outlined in the previous subsections. On closer inspection, however, we see that the character of the problem has changed dramatically, which might occasionally lead to qualitatively new effects. This becomes immediately clear once we perform a Taylor expansion of the dielectric tensor into powers of the inverse wavelength. To first order, we get
\begin{align}
 \tsr \varepsilon_{\rm eff}(\hat{\vec k},|\vec k|,\omega) 
 &\approx\tsr\varepsilon_{\rm eff}(\hat{\vec k},\omega) + \frac{\omega}{c}n(\hat{\vec k},\omega)\left(\frac{\partial}{\partial|\vec k|}\tsr \varepsilon_{\rm eff}\right)_{|\vec k|=0}(\hat{\vec k},\omega) \,,
\end{align}
and this can be fed back into Eq.~\eqref{det_condmod}, where the new term completely absent in the preceding subsections would spoil the entire formulary. For the sake of simplicity, we illustrate this on the most simple example, namely on our starting point, the standard relation between the refractive index and the (longitudinal) dielectric function, Eq.~\eqref{eq_standard}, to which the above formula reduces itself under the crassest simplifying assumptions. To first order Taylor expansion, this standard formula would now have to be replaced by
\begin{equation}
 \varepsilon(\omega) + \varepsilon'(\omega)\frac{\omega}{c}n(\omega) = n^2(\omega) \,,
\end{equation}
where the prime denotes the derivative w.r.t.~the modulus of the wavevector.
And this can be solved for the refractive indices as
\begin{equation}
 n_{1/2}(\omega) = \frac 1 2 \varepsilon'(\omega)\frac{\omega}{c} \pm \sqrt{\left(\frac 1 2\varepsilon'(\omega)\frac{\omega}{c}\right)^2 + \varepsilon(\omega)} \,. \label{eq_corrections}
\end{equation}
The upshot of this is that for wavelength dependent dielectric functions, one may now have two essentially different refractive indices even in the completely isotropic case where the longitudinal and transverse proper conductivities coincide. Even worse---or rather even better, if one is only interested in new physics---, if the Taylor expansion was continued to arbitrarily high orders, the determining equation \eqref{det_condmod} for the refractive index would correspond to the zeros of a higher order polynomial and could therefore lead to much more essentially different refractive indices at the same frequency and direction than the usual two in the tensorial though wavelength independent case. As this, however, does not seem to apply in nature, we can safely discard the possibility that in the optical range the effective dielectric tensor would depend on the wavelength for ``ordinary'' materials. 

Apart from the hypothetical chance that the effective dielectric tensor depends on the wavelength for some extremely specific material nonetheless, the benefit of this consideration is that we finally obtain a certain clarity about the frequently invoked optical limit $|\vec k|\rightarrow 0$, which is allegedly to be performed before optical properties can be accessed from ab initio outputs. In fact, in this respect we can say this: Either the effective dielectric tensor is independent of the wavelength anyway---in which case this limit is quite superfluous and the effective dielectric tensor can simply be calculated for infinite wavelengths at the very outset---or it is not. If it is not, however, this limit will be plainly wrong, no matter how unlikely this case actually is. In fact, it makes no sense to apply such a limit to the fundamental equation \eqref{implicit} and correspondingly in the presence of a perceptible wavelength dependence in the effective dielectric tensor we necessarily get higher order correction terms instead, which can then not be neglected (because this would precisely mean that the wavelength dependence can be neglected all the same).

\section{Applications} \label{appendix}

For the convenience of the reader, we reproduce in this final section some standard results 
from the fundamental formulae \eqref{eq_central5} and \eqref{exact_sol} by introducing suitable approximations.

\subsection{Vacuum} \label{app:vacuum}
First, in the absence of a medium, the conductivity tensor vanishes and hence the effective dielectric tensor equals the identity matrix. 
Thus, in vacuo, the only solution of Eq.~\eqref{det_mat} is $n^2 = 1$ corresponding to the dispersion relation $\omega_{\vec k} = c|\vec k|$. 
Furthermore, for this refractive index there are two orthogonal polarization vectors $\vec e_{\vec k1}$ and $\vec e_{\vec k2}$ which are both purely transverse.

\subsection{Isotropic material} \label{subsec:iso}

Next, for an isotropic material, we have only two independent components of the (effective) dielectric tensor\footnote{In fact,
for a ``strictly'' isotropic material, the respective dielectric functions would not even depend on the direction of the wavevector but at best
on its modulus, which is, however, excluded in the present context (see the discussion in Ref.~\cite[Appendix D.1]{EffWW}). We also note that, either way, isotropic response functions necessarily depend on the direction of the wavevector. Thus, a completely wavevector independent conductivity tensor can never be isotropic except for the case where it is proportional to the identity matrix.}:
\begin{align}
 \varepsilon_{11}(\hat{\vec k}, \omega) = \varepsilon_{22}(\hat{\vec k}, \omega) & = \varepsilon_{\rm eff,\hh T}(\hat{\vec k}, \omega) \,, \\[3pt]
 \varepsilon_{33}(\hat{\vec k}, \omega) & = \varepsilon_{\rm eff, \hh L}(\hat{\vec k}, \omega) \,.
\end{align}
Correspondingly, the refractive indices are determined by the condition
\begin{equation} \label{iso_cond}
 \left(\varepsilon_{\rm eff, \hh T}(\hat{\vec k}, \omega) - n^2(\hat{\vec k}, \omega) \right)^{\!2} \varepsilon_{\rm eff, \hh L}(\hat{\vec k}, \omega) = 0 \,.
\end{equation}
We further distinguish between the following two cases. First, we consider $\varepsilon_{\rm eff, \hh L}(\hat{\vec k}, \omega) = 0$. 
This condition determines the frequency as a function of the direction. In fact, those frequencies $\omega = \omega(\hat{\vec k})$ for which the longitudinal \linebreak dielectric function vanishes 
are precisely the {\itshape plasmon frequencies} in the iso\-{}tropic case (see Eq.~\eqref{eq_long}). The corresponding polarization vectors are purely longitudinal. By contrast, the concept of a refractive index is not meaningful in this case because Eq.~\eqref{iso_cond} is then fulfilled for any $n$.
Secondly,~we consider $\varepsilon_{\rm eff, \hh L}(\hat{\vec k}, \omega) \not = 0$. Away from the plasmon frequencies, Eq.~\eqref{iso_cond} has a two-fold root given in terms of the transverse dielectric function by
\begin{equation} \label{cite_this}
 n^2(\hat{\vec k}, \omega) = \varepsilon_{\rm eff,T}(\hat{\vec k}, \omega) \,.
\end{equation}
We note that this equation referring to the {\itshape effective} transverse dielectric function is fully equivalent to its counterpart Eq.~\eqref{eq_trans_disp}, which is formulated in terms of the fundamental ({\itshape ab initio}) transverse dielectric function. Furthermore, in this second case, any two transverse, mutually orthogonal vectors $\vec e_{\vec k 1}$ and $\vec e_{\vec k 2}$ can be regarded as polarization vectors, which share the same refractive index given by Eq.~\eqref{cite_this}. Thus, for an isotropic medium we recover the well-known results described already in \S\,\ref{Subsct:genwave}. 

\subsection{Optical activity}

A more general case, which includes the isotropic limit, is defined by the absence of the longitudinal-transverse cross-couplings, i.e., by
\begin{equation}
 \tsr P\L(\hat{\vec k}) \, \tsr\varepsilon_{\rm eff}(\omega) \, \tsr P\T(\hat{\vec k}) = \tsr P\T(\hat{\vec k}) \,\tsr\varepsilon_{\rm eff}(\omega) \, \tsr P\L(\vec k) = 0\,.
\end{equation}
In this case, the fundamental Eq.~\eqref{det_mat} simplifies to
\begin{equation} \label{det_mat1}
 \det \mh \left( \begin{array}{lll} \varepsilon_{11} - n^2 & \varepsilon_{12} & 0 \\[3pt] \varepsilon_{21} & \varepsilon_{22} - n^2 & 0 
 \\[3pt] 0 & 0 & \varepsilon_{33} \end{array} \right) = 0 \,.
 \end{equation}
Now, the generalized plasmon equation \eqref{eq_plasmon} together with the condition $\vec e\T(\hat{\vec k}, \omega) = 0$ 
simply implies $\varepsilon_{33}(\hat{\vec k}, \omega)=0$, and the corresponding frequencies~are precisely the plasmon frequencies. 
Furthermore, there exist at most two {\itshape purely transverse} polarization vectors corresponding
to two, in general different, refractive indices. These can be characterized as the eigenvectors and eigenvalues
of the $(2\times2)$ block matrix corresponding to the transverse subspace. Again, this block matrix does in general not have to be hermitean.
It may, in special cases, be of the form $\varepsilon_{11}=\varepsilon_{22}$ and $\varepsilon_{12}=-\varepsilon_{21}$ 
(which, in fact, would be hermitean if only the diagonal entries were real and the off-diagonal entries were purely imaginary), 
such that the general formula \eqref{exact_sol} implies $n_\pm^2 = \varepsilon_{11} \pm \i \h \varepsilon_{12}$ 
and the eigenvectors turn out to be of the form $\vec e_\pm = (\vec e_1 \pm \i \h \vec e_2)/\sqrt{2}$ (see Ref.~\cite[p.~182)]{Lipson}). 
In this case, the dielectric tensor induces {\itshape optical activity} in the respective direction (see e.g.~Ref.~\cite[Eq.~(6.39)]{Lipson}).
Within the realm of solid state physics, the arch example for this~type of behavior is provided by $\alpha$-quartz.
On the other hand, for \linebreak $\varepsilon_{11} = \varepsilon_{22}$ and $\varepsilon_{12} = \varepsilon_{21} = 0$, we recover again the results from the previous subsection.

\subsection{Birefringence}

As a matter of principle, the formalism presented in this article encompasses birefringence as well.
Experimentally, a typical example for this type of behavior is given by rutile (TiO$_2$). 
For practical applications in the optical industry, such materials as $\alpha$-BBO (Ba(BO$_2$)$_2$), LiNbO$_3$, or LiTaO$_3$ are relevant examples.
To demonstrate that birefringence can be described in our formalism, we consider here for the sake of simplicity the case of a {\itshape uniaxial birefringent material,} whose 
proper conductivity tensor is of the form
\begin{equation}
 \tsr{\widetilde\sigma}(\omega) = \sigma_0(\omega) \h \tsr{1} + (\Delta\sigma)(\omega) \, \hat{\vec a} \hh \hat{\vec a}^{\rm T} \,,
\end{equation}
where $\hat{\vec a}$ denotes a fixed unit vector in {\itshape real space}, the so-called {\itshape optical axis}. In this case,
all direction-dependent refractive indices can be calculated analytically from our formalism. For this purpose, let us define the reference indices
\begin{align}
 n_1^2(\omega) &= 1 - \frac{\sigma_0(\omega)}{\i\omega\h \varepsilon_0} \,, \\[5pt]
 n_2^2(\omega) &= 1 - \frac{\sigma_0(\omega)+(\Delta\sigma)(\omega)}{\i\omega\h \varepsilon_0} \,.
\end{align}
In terms of these, the {\itshape effective} dielectric tensor can be written as
\begin{equation}
 \tsr\varepsilon_{\rm eff}(\omega) \h \equiv \h \tsr 1 - \frac{\tsr{\widetilde \sigma}(\omega)}{\j\omega \h \varepsilon_0}  \h = \h n_1^2(\omega) \h \tsr P_{\rm T}(\hat{\vec a}) + n_2^2(\omega) \h \tsr P_{\rm L}(\hat{\vec a}) \,,
\end{equation}
where $P_{\rm L}(\hat{\vec a})$ and $P_{\rm T}(\hat{\vec a})$ denote the longitudinal and transverse projection operators 
{\it in the direction of} the optical axis $\hat{\vec a}$. We first note that for \mbox{$\hat{\vec k} = \hat{\vec a}$}, i.e., if the wavevector is parallel to the optical axis, we recover the results derived in the isotropic case (see \S\,\ref{subsec:iso}), with two purely transverse polarization vectors sharing the same refractive index $n_1^2(\omega)$. The situation is more complicated in the case where $\hat{\vec k} \not = \hat{\vec a}$. Then, we make the ansatz
\begin{equation}
 \vec e_{\rm or}(\hat{\vec k}) = \frac{\hat{\vec k} \times \hat{\vec a}}{|\hat{\vec k} \times \hat{\vec a}|}
\end{equation}
for the first polarization vector, which is both orthogonal to the wavevector (i.e., purely transverse) and to the optical axis. One sees directly 
that this polarization vector indeed solves the central equation \eqref{eq_central} with the refractive
index $n_{\rm or}^2(\omega) = n_1^2(\omega)$. Hence, the first refractive index is independent of the direction although the polarization vector is not.
This solution corresponds to the so-called {\itshape ordinary ray}. In addition, one has to consider the {\itshape extraordinary ray},
for whose polarization vector we  make the ansatz
\begin{equation}
 \vec e_{\rm ex}(\hat{\vec k}, \omega) = A(\hat{\vec k}, \omega) \, \hat{\vec k} + B(\hat{\vec k}, \omega) \, \hat{\vec a} \,,
\end{equation}
such that $\vec e_{\rm ex}$ is perpendicular to $\vec e_{\rm or}$\h. The two scalar functions $A$ and $B$ have to be determined together with the refractive index $n_{\rm ex}(\hat{\vec k}, \omega)$ by putting this ansatz into the central equation \eqref{eq_central} 
and taking into account the normalization condition $|\vec e_{\rm ex}(\hat{\vec k}, \omega)| = 1$. Defining the angle $\alpha(\hat{\vec k})$ between $\hat{\vec k}$~and~$\hat{\vec a}$~by
\begin{equation}
 \cos\hh[\alpha(\hat{\vec k})] = \hat{\vec k} \cdot \hat{\vec a} \,,
\end{equation}
we thus obtain after a lengthy but straightforward calculation the refractive index of the extraordinary ray,
\begin{equation}
 n_{\rm ex}^2 = \frac{n^2_1 \, n^2_2}{n_1^2\sin^2\alpha + n_2^2\cos^2\alpha} \,,
\end{equation}
as well as the coefficient functions
\begin{equation}
 B = \bigg(\sin^2 \alpha + \bigg( 1 - \frac{n_{\rm ex}^2}{n_1^2} \bigg)^{\!\!2} \cos^2 \alpha \h \bigg)^{\!\!-1/2} \,,
\end{equation}
and \smallskip
\begin{equation}
 A = -\frac{n_{\rm ex}^2 \cos\alpha}{n_1^2} \, B \,. \smallskip
\end{equation}
In particular, we see that the polarization vector of the extraordinary ray is not purely transverse but encloses an angle $\varphi_{\rm ex}$ with the wavevector given by
\begin{equation}
 \cos \varphi_{\rm ex} \h \equiv \h \hat{\vec k} \cdot \vec e_{\rm ex} \h = \h A + B \hh \cos\alpha \,.
\end{equation}
One can show that in this case the ray direction (which is---in general---defined by the Poynting vector) 
and the direction of the wavevector do not coincide---fact which accounts for the ``extraordinary'' behaviour. 
Finally, for $\alpha \to 0$, one sees that $n_{\rm ex}^2 \to n_1^2$ and \mbox{$(\cos \varphi_{\rm ex}) \to 0$,} thus the extraordinary ray becomes purely transverse and shares the same refractive index with the ordinary ray, which is consistent with the results discussed in the case where $\hat{\vec k} = \hat{\vec a}$. All of this is in complete accordance with the standard results for birefringent materials (see e.g.~Ref.~\cite[\S\,4.5]{Bergmann}).

\section{Conclusion}

%

We have concisely criticized the standard calculation of the refractive index from the {\itshape scalar,} wavevector-{\itshape independent} dielectric function.
Correspondingly, we have based our treatment of the refractive index on the fundamental, microscopic wave equation in materials, Eq.~\eqref{eq_general}, which involves in general a wavevector-dependent dielectric tensor. We have shown that this approach even unifies the equations for optical waves and plasmons. Generally, our central results are these:
\begin{enumerate}
\item We have proven the equivalence of this fundamental, microscopic wave equation to the standard wave equation used in theoretical optics, Eq.~\eqref{eq_transWaveEqMod}, under the assumption that the latter refers to the {\itshape effective} dielectric tensor \eqref{eq_defEffDiTens} 
rather than to the fundamental ({\itshape ab initio}) dielectric tensor \eqref{eq_fundDef}. 
\item Thereby, we have shown that  the combination of {\itshape ab initio} methods for calculating the proper conductivity tensor---which is a standard target quantity of any modern {\itshape ab initio} materials simulation code \cite{elk, VASP, Wien2k, QE-2009} (see also the discussion in Ref.~\cite[\S\,II]{Schwalbe})---with the Fresnel equation from theoretical optics, Eq.~\eqref{poly}, solves the problem of calculating wavevector-dependent optical properties from wavevector-independent
response functions. 
\item Correspondingly, the central equation for the joint determination of frequency- and direction-dependent refractive indices and their respective polarization vectors is given by Eq.~\eqref{eq_central}.
\item We have clarified that the application of this entire formalism only requires the wavelength independence of the proper conductivity tensor, while a direction dependence is still possible, apart from the frequency dependence which is always assumed to be present (see Sct.~\ref{Sct:OpticalProp}).
\item We have also investigated the consequences of a hypothetical wavelength dependence in the proper conductivity tensor. In particular, we have shown that the fundamental determining equation for both refractive indices and polarization vectors, Eq.~\eqref{det_cond}, can then still be turned into the closed though implicit equation \eqref{det_condmodmod}, which leads directly to correction terms as in Eq.~\eqref{eq_corrections} via a Taylor expansion in the modulus of the wavevector.
\item In particular, this makes it clear that the so-called optical limit $|\vec k|\rightarrow 0$ as performed in the dielectric tensor before calculating refractive indices is either superfluous (because the effective dielectric tensor is practically wavelength independent anyway) or wrong (because it illegitimately neglects correction terms which would be relevant). As these correction terms probably lead to qualitatively new effects, which could be detected independently of any prior ab initio calculation, we also obtain the not entirely trivial {\it empirical} conclusion that the proper conductivity tensor of typical materials does indeed not depend on the wavelength in the optical range.
\end{enumerate}

Besides these results, we have also clarified some more general theoretical issues such as the following:
(i) The standard formula \eqref{eq_standard} for the refractive index actually only holds under the condition of coinciding longitudinal and transverse proper conductivities (see Eq.~\eqref{eq_longAppr}).
(ii) If this condition is not fulfilled, it should be replaced by Eq.~\eqref{implicit} combined with Eq.~\eqref{eq_defRefrInd}. 
(iii) In the standard treatment, the assumption of wavelength independence actually applies to the proper conductivity tensor (see Eq.~\eqref{cond_cond}).
(vi) The relation between the dielectric tensors \eqref{eq_fundDef} and \eqref{eq_defEffDiTens} 
used respectively in {\itshape ab initio} physics and theoretical optics is given by Eq.~\eqref{cond_rel_2}).

\section*{Acknowledgments}
This research was supported by the DFG grant HO 2422/12-1 and by the DFG RTG 1995. R.\,S. thanks the Institute for Theoretical Physics at TU Bergakademie Freiberg for its hospitality. We further thank Heinrich-Gregor Zirnstein for feedback on the manuscript.

\bigskip
\bibliographystyle{model1-num-names.bst}
\bibliography{masterbib}

\begin{thebibliography}{59}
\expandafter\ifx\csname natexlab\endcsname\relax\def\natexlab#1{#1}\fi
\providecommand{\url}[1]{\texttt{#1}}
\providecommand{\href}[2]{#2}
\providecommand{\path}[1]{#1}
\providecommand{\DOIprefix}{doi:}
\providecommand{\ArXivprefix}{arXiv:}
\providecommand{\URLprefix}{URL: }
\providecommand{\Pubmedprefix}{pmid:}
\providecommand{\doi}[1]{\href{http://dx.doi.org/#1}{\path{#1}}}
\providecommand{\Pubmed}[1]{\href{pmid:#1}{\path{#1}}}
\providecommand{\bibinfo}[2]{#2}
\ifx\xfnm\relax \def\xfnm[#1]{\unskip,\space#1}\fi
\bibitem[{Schober and Starke(2018)}]{EDWegner}
\bibinfo{author}{G.~A.~H. Schober}, \bibinfo{author}{R.~Starke},
\newblock \bibinfo{title}{{\itshape Microscopic theory of refractive index
  applied to metamaterials: effective current response tensor corresponding to
  standard relation $n^2 = \varepsilon_{\textnormal{eff}} \hspace{1pt}
  \mu_{\textnormal{eff}}$}},
\newblock \bibinfo{journal}{Eur. Phys. J. B} \bibinfo{volume}{{\bfseries 91}}
  (\bibinfo{year}{2018}) \bibinfo{pages}{146}. \bibinfo{note}{{See also
  arXiv:1709.08811 [physics.class-ph]}}.
\bibitem[{Strinati(1988)}]{Strinati}
\bibinfo{author}{G.~Strinati},
\newblock \bibinfo{title}{{\itshape Application of the Green's functions method
  to the study of optical properties of semiconductors}},
\newblock \bibinfo{journal}{La Rivista del Nuovo Cimento}
  \bibinfo{volume}{{\bfseries 11}} (\bibinfo{year}{1988}) \bibinfo{pages}{1}.
\bibitem[{Hanke(1978)}]{Hanke}
\bibinfo{author}{W.~Hanke},
\newblock \bibinfo{title}{{\itshape Dielectric theory of elementary excitations
  in crystals}},
\newblock \bibinfo{journal}{Adv. Phys.} \bibinfo{volume}{{\bfseries 27}}
  (\bibinfo{year}{1978}) \bibinfo{pages}{287}.
\bibitem[{Bechstedt(2015)}]{Bechstedt}
\bibinfo{author}{F.~Bechstedt}, \bibinfo{title}{{\itshape Many-body approach to
  electronic excitations: concepts and applications}}, volume
  \bibinfo{volume}{181} of \textit{\bibinfo{series}{\textnormal{Springer Series
  in Solid-State Sciences}}}, \bibinfo{publisher}{Springer-Verlag},
  \bibinfo{address}{Berlin/Heidelberg}, \bibinfo{year}{2015}.
\bibitem[{Yu and Cardona(2010)}]{Cardona}
\bibinfo{author}{P.~Y. Yu}, \bibinfo{author}{M.~Cardona},
  \bibinfo{title}{{\itshape Fundamentals of semiconductors: physics and
  materials properties}}, \textnormal{Graduate Texts in Physics},
  \bibinfo{edition}{4th} ed., \bibinfo{publisher}{Springer-Verlag},
  \bibinfo{address}{Berlin/Heidelberg}, \bibinfo{year}{2010}.
\bibitem[{Ashcroft and Mermin(1976)}]{Ashcroft}
\bibinfo{author}{N.~W. Ashcroft}, \bibinfo{author}{N.~D. Mermin},
  \bibinfo{title}{{\itshape Solid state physics}},
  \bibinfo{publisher}{Harcourt, Inc.}, \bibinfo{address}{Orlando, FL},
  \bibinfo{year}{1976}.
\bibitem[{Sch\"afer and Wegener(2002)}]{SchafWegener}
\bibinfo{author}{W.~Sch\"afer}, \bibinfo{author}{M.~Wegener},
  \bibinfo{title}{{\itshape Semiconductor optics and transport phenomena}},
  \textnormal{Advanced Texts in Physics}, \bibinfo{publisher}{Springer-Verlag},
  \bibinfo{address}{Berlin/Heidelberg}, \bibinfo{year}{2002}.
\bibitem[{Pines and Nozi\`{e}res(1999)}]{NozieresPinesBook}
\bibinfo{author}{D.~Pines}, \bibinfo{author}{P.~Nozi\`{e}res},
  \bibinfo{title}{{\itshape The theory of quantum liquids}},
  \textnormal{Advanced Book Classics}, \bibinfo{publisher}{Perseus Books
  Publishing, L.\,L.\,C.}, \bibinfo{address}{Cambridge, MA},
  \bibinfo{year}{1999}.
\bibitem[{Berger et~al.(2007)Berger, de~Boeij, and van Leeuwen}]{Berger}
\bibinfo{author}{J.~A. Berger}, \bibinfo{author}{P.~L. de~Boeij},
  \bibinfo{author}{R.~van Leeuwen},
\newblock \bibinfo{title}{{\itshape Analysis of the {Vignale-Kohn} current
  functional in the calculation of the optical spectra of semiconductors}},
\newblock \bibinfo{journal}{Phys. Rev. B} \bibinfo{volume}{{\bfseries 75}}
  (\bibinfo{year}{2007}) \bibinfo{pages}{035116}.
\bibitem[{Lee et~al.(2011)Lee, Schober, Bahramy, Murakawa, Onose, Arita,
  Nagaosa, and Tokura}]{Lee}
\bibinfo{author}{J.~S. Lee}, \bibinfo{author}{G.~A.~H. Schober},
  \bibinfo{author}{M.~S. Bahramy}, \bibinfo{author}{H.~Murakawa},
  \bibinfo{author}{Y.~Onose}, \bibinfo{author}{R.~Arita},
  \bibinfo{author}{N.~Nagaosa}, \bibinfo{author}{Y.~Tokura},
\newblock \bibinfo{title}{{\itshape Optical response of relativistic electrons
  in the polar $\mathrm{BiTeI}$ semiconductor}},
\newblock \bibinfo{journal}{Phys. Rev. Lett.} \bibinfo{volume}{{\bfseries 107}}
  (\bibinfo{year}{2011}) \bibinfo{pages}{117401}.
\bibitem[{Demk\'o et~al.(2012)Demk\'o, Schober, Kocsis, Bahramy, Murakawa, Lee,
  K\'ezsm\'arki, Arita, Nagaosa, and Tokura}]{Laszlo}
\bibinfo{author}{L.~Demk\'o}, \bibinfo{author}{G.~A.~H. Schober},
  \bibinfo{author}{V.~Kocsis}, \bibinfo{author}{M.~S. Bahramy},
  \bibinfo{author}{H.~Murakawa}, \bibinfo{author}{J.~S. Lee},
  \bibinfo{author}{I.~K\'ezsm\'arki}, \bibinfo{author}{R.~Arita},
  \bibinfo{author}{N.~Nagaosa}, \bibinfo{author}{Y.~Tokura},
\newblock \bibinfo{title}{{\itshape Enhanced infrared magneto-optical response
  of the nonmagnetic semiconductor $\mathrm{BiTeI}$ driven by bulk Rashba
  splitting}},
\newblock \bibinfo{journal}{Phys. Rev. Lett.} \bibinfo{volume}{{\bfseries 109}}
  (\bibinfo{year}{2012}) \bibinfo{pages}{167401}.
\bibitem[{Kawai et~al.(2014)Kawai, Yamashita, Cannuccia, and Marini}]{Kawai}
\bibinfo{author}{H.~Kawai}, \bibinfo{author}{K.~Yamashita},
  \bibinfo{author}{E.~Cannuccia}, \bibinfo{author}{A.~Marini},
\newblock \bibinfo{title}{{\itshape Electron-electron and electron-phonon
  correlation effects on the finite-temperature electronic and optical
  properties of zinc-blende $\mathrm{GaN}$}},
\newblock \bibinfo{journal}{Phys. Rev. B} \bibinfo{volume}{{\bfseries 89}}
  (\bibinfo{year}{2014}) \bibinfo{pages}{085202}.
\bibitem[{Makhnev et~al.(2014)Makhnev, Nomerovannaya, Kuznetsova, Tereshchenko,
  and Kokh}]{Makhnev}
\bibinfo{author}{A.~A. Makhnev}, \bibinfo{author}{L.~V. Nomerovannaya},
  \bibinfo{author}{T.~V. Kuznetsova}, \bibinfo{author}{O.~E. Tereshchenko},
  \bibinfo{author}{K.~A. Kokh},
\newblock \bibinfo{title}{{\itshape Optical properties of $\mathrm{BiTeI}$
  semiconductor with a strong Rashba spin-orbit interaction}},
\newblock \bibinfo{journal}{Opt. Spectrosc.} \bibinfo{volume}{{\bfseries 117}}
  (\bibinfo{year}{2014}) \bibinfo{pages}{764}.
\bibitem[{Rusinov et~al.(2015)Rusinov, Tereshchenko, Kokh, Shakhmametova,
  Azarov, and Chulkov}]{Rusinov15}
\bibinfo{author}{I.~P. Rusinov}, \bibinfo{author}{O.~E. Tereshchenko},
  \bibinfo{author}{K.~A. Kokh}, \bibinfo{author}{A.~R. Shakhmametova},
  \bibinfo{author}{I.~A. Azarov}, \bibinfo{author}{E.~V. Chulkov},
\newblock \bibinfo{title}{{\itshape Role of anisotropy and spin-orbit
  interaction in the optical and dielectric properties of $\mathrm{BiTeI}$ and
  $\mathrm{BiTeCl}$ compounds}},
\newblock \bibinfo{journal}{JETP Lett.} \bibinfo{volume}{{\bfseries 101}}
  (\bibinfo{year}{2015}) \bibinfo{pages}{507}. \bibinfo{note}{[Pis'ma Zh. Eksp.
  Teor. Fiz. {\bfseries 101}, 563 (2015)]}.
\bibitem[{Akrap et~al.(2014)Akrap, Teyssier, Magrez, Bugnon, Berger, Kuzmenko,
  and van~der Marel}]{Akrap}
\bibinfo{author}{A.~Akrap}, \bibinfo{author}{J.~Teyssier},
  \bibinfo{author}{A.~Magrez}, \bibinfo{author}{P.~Bugnon},
  \bibinfo{author}{H.~Berger}, \bibinfo{author}{A.~B. Kuzmenko},
  \bibinfo{author}{D.~van~der Marel},
\newblock \bibinfo{title}{{\itshape Optical properties of $\mathrm{BiTeBr}$ and
  $\mathrm{BiTeCl}$}},
\newblock \bibinfo{journal}{Phys. Rev. B} \bibinfo{volume}{{\bfseries 90}}
  (\bibinfo{year}{2014}) \bibinfo{pages}{035201}.
\bibitem[{Dongho~Nguimdo and Joubert(2015)}]{dongho_nguimdo_density_2015}
\bibinfo{author}{G.~M. Dongho~Nguimdo}, \bibinfo{author}{D.~P. Joubert},
\newblock \bibinfo{title}{{\itshape A density functional (PBE, PBEsol, HSE06)
  study of the structural, electronic and optical properties of the ternary
  compounds {AgAlX}2 (X = S, Se, Te)}},
\newblock \bibinfo{journal}{Eur. Phys. J. B} \bibinfo{volume}{{\bfseries 88}}
  (\bibinfo{year}{2015}) \bibinfo{pages}{113}.
\bibitem[{Gracia et~al.(2009)Gracia, Beltran, and
  Errandonea}]{gracia_characterization_2009}
\bibinfo{author}{L.~Gracia}, \bibinfo{author}{A.~Beltran},
  \bibinfo{author}{D.~Errandonea},
\newblock \bibinfo{title}{{\itshape Characterization of the TiSiO4 structure
  and its pressure-induced phase transformations: Density functional theory
  study}},
\newblock \bibinfo{journal}{Phys. Rev. B} \bibinfo{volume}{{\bfseries 80}}
  (\bibinfo{year}{2009}).
\bibitem[{L\"oper et~al.(2015)L\"oper, Stuckelberger, Niesen, Werner, Filipic,
  Moon, Yum, Topic, De~Wolf, and Ballif}]{loper_complex_2015}
\bibinfo{author}{P.~L\"oper}, \bibinfo{author}{M.~Stuckelberger},
  \bibinfo{author}{B.~Niesen}, \bibinfo{author}{J.~Werner},
  \bibinfo{author}{M.~Filipic}, \bibinfo{author}{S.-J. Moon},
  \bibinfo{author}{J.-H. Yum}, \bibinfo{author}{M.~Topic},
  \bibinfo{author}{S.~De~Wolf}, \bibinfo{author}{C.~Ballif},
\newblock \bibinfo{title}{{\itshape Complex Refractive Index Spectra of $CH_3
  NH_3 PbI_3 $ Perovskite Thin Films Determined by Spectroscopic Ellipsometry
  and Spectrophotometry}},
\newblock \bibinfo{journal}{J. Phys. Chem. Lett.} \bibinfo{volume}{{\bfseries
  6}} (\bibinfo{year}{2015}) \bibinfo{pages}{66--71}.
\bibitem[{Saha et~al.(2000)Saha, Sinha, and Mookerjee}]{saha_electronic_2000}
\bibinfo{author}{S.~Saha}, \bibinfo{author}{T.~P. Sinha},
  \bibinfo{author}{A.~Mookerjee},
\newblock \bibinfo{title}{{\itshape Electronic structure, chemical bonding, and
  optical properties of paraelectric BaTiO3}},
\newblock \bibinfo{journal}{Phys. Rev. B} \bibinfo{volume}{{\bfseries 62}}
  (\bibinfo{year}{2000}) \bibinfo{pages}{8828--8834}.
\bibitem[{Friedrich et~al.(2017)Friedrich, Schmidt, Schindlmayr, and
  Sanna}]{friedrich_optical_2017}
\bibinfo{author}{M.~Friedrich}, \bibinfo{author}{W.~G. Schmidt},
  \bibinfo{author}{A.~Schindlmayr}, \bibinfo{author}{S.~Sanna},
\newblock \bibinfo{title}{{\itshape Optical properties of titanium-doped
  lithium niobate from time-dependent density-functional theory}},
\newblock \bibinfo{journal}{Phys. Rev. Materials} \bibinfo{volume}{{\bfseries
  1}} (\bibinfo{year}{2017}).
\bibitem[{Yamada et~al.(2018)Yamada, Noda, Nobusada, and
  Yabana}]{yamada_time-dependent_2018}
\bibinfo{author}{S.~Yamada}, \bibinfo{author}{M.~Noda},
  \bibinfo{author}{K.~Nobusada}, \bibinfo{author}{K.~Yabana},
\newblock \bibinfo{title}{{\itshape Time-dependent density functional theory
  for interaction of ultrashort light pulse with thin materials}},
\newblock \bibinfo{journal}{Phys. Rev. B} \bibinfo{volume}{{\bfseries 98}}
  (\bibinfo{year}{2018}).
\bibitem[{Sangalli et~al.(2017)Sangalli, Berger, Attaccalite, Gr\"{u}ning, and
  Romaniello}]{Myrta}
\bibinfo{author}{D.~Sangalli}, \bibinfo{author}{J.~A. Berger},
  \bibinfo{author}{C.~Attaccalite}, \bibinfo{author}{M.~Gr\"{u}ning},
  \bibinfo{author}{P.~Romaniello},
\newblock \bibinfo{title}{{\itshape Optical properties of periodic systems
  within the current-current response framework: pitfalls and remedies}},
\newblock \bibinfo{journal}{Phys.~Rev.~B} \bibinfo{volume}{{\bfseries 95}}
  (\bibinfo{year}{2017}) \bibinfo{pages}{155203}.
\bibitem[{Forcella et~al.(2017)Forcella, Prada, and Carminati}]{Forcella}
\bibinfo{author}{D.~Forcella}, \bibinfo{author}{C.~Prada},
  \bibinfo{author}{R.~Carminati},
\newblock \bibinfo{title}{{\itshape Causality, Nonlocality, and Negative
  Refraction}},
\newblock \bibinfo{journal}{Phys. Rev. Lett.} \bibinfo{volume}{{\bfseries 118}}
  (\bibinfo{year}{2017}) \bibinfo{pages}{134301}.
\bibitem[{Del~Sole and Fiorino(1984)}]{Fiorino}
\bibinfo{author}{R.~Del~Sole}, \bibinfo{author}{E.~Fiorino},
\newblock \bibinfo{title}{{\itshape Macroscopic dielectric tensor at crystal
  surfaces}},
\newblock \bibinfo{journal}{Phys. Rev. B} \bibinfo{volume}{{\bfseries 29}}
  (\bibinfo{year}{1984}) \bibinfo{pages}{4631}.
\bibitem[{Starke and Schober(2015)}]{ED1}
\bibinfo{author}{R.~Starke}, \bibinfo{author}{G.~A.~H. Schober},
\newblock \bibinfo{title}{{\itshape Functional Approach to electrodynamics of
  media}},
\newblock \bibinfo{journal}{Phot. Nano. Fund. Appl.}
  \bibinfo{volume}{{\bfseries 14}} (\bibinfo{year}{2015})
  \bibinfo{pages}{1--34}. \bibinfo{note}{{See also arXiv:1401.6800
  [cond-mat.mtrl-sci]}}.
\bibitem[{Starke and Schober(2016)}]{ED2}
\bibinfo{author}{R.~Starke}, \bibinfo{author}{G.~A.~H. Schober},
  \bibinfo{title}{{\itshape Ab initio materials physics and microscopic
  electrodynamics of media}}, \bibinfo{howpublished}{arXiv:1606.00445
  [cond-mat.mtrl-sci]}, \bibinfo{year}{2016}.
\bibitem[{Starke and Schober(2017)}]{Refr}
\bibinfo{author}{R.~Starke}, \bibinfo{author}{G.~A.~H. Schober},
\newblock \bibinfo{title}{{\itshape Microscopic theory of the refractive
  index}},
\newblock \bibinfo{journal}{Optik} \bibinfo{volume}{{\bfseries 140}}
  (\bibinfo{year}{2017}) \bibinfo{pages}{62}. \bibinfo{note}{{See also
  arXiv:1510.03404 [cond-mat.mtrl-sci]}}.
\bibitem[{Giuliani and Vignale(2005)}]{Giuliani}
\bibinfo{author}{G.~F. Giuliani}, \bibinfo{author}{G.~Vignale},
  \bibinfo{title}{{\itshape Quantum theory of the electron liquid}},
  \bibinfo{publisher}{Cambridge University Press},
  \bibinfo{address}{Cambridge}, \bibinfo{year}{2005}.
\bibitem[{Altland and Simons(2010)}]{Altland}
\bibinfo{author}{A.~Altland}, \bibinfo{author}{B.~Simons},
  \bibinfo{title}{{\itshape Condensed matter field theory}},
  \bibinfo{edition}{2nd} ed., \bibinfo{publisher}{Cambridge University Press},
  \bibinfo{address}{Cambridge}, \bibinfo{year}{2010}.
\bibitem[{Melrose(2008)}]{Melrose1Book}
\bibinfo{author}{D.~B. Melrose}, \bibinfo{title}{{\itshape Quantum
  plasmadynamics: unmagnetized plasmas}}, volume \bibinfo{volume}{735} of
  \textit{\bibinfo{series}{\textnormal{Lecture Notes in Physics}}},
  \bibinfo{publisher}{Springer}, \bibinfo{address}{New York},
  \bibinfo{year}{2008}.
\bibitem[{elk(enet)}]{elk}
\bibinfo{title}{{Elk FP-LAPW Code}},
  \bibinfo{year}{\url{http://elk.sourceforge.net}}.
\bibitem[{Starke and Schober(2016{\natexlab{a}})}]{EDOhm}
\bibinfo{author}{R.~Starke}, \bibinfo{author}{G.~A.~H. Schober},
\newblock \bibinfo{title}{{\itshape Relativistic covariance of {Ohm's} law}},
\newblock \bibinfo{journal}{Int. J. Mod. Phys. D} \bibinfo{volume}{{\bfseries
  25}} (\bibinfo{year}{2016}{\natexlab{a}}) \bibinfo{pages}{1640010}.
  \bibinfo{note}{{See also arXiv:1409.3723 [math-ph]}}.
\bibitem[{Starke and Schober(2016{\natexlab{b}})}]{EffWW}
\bibinfo{author}{R.~Starke}, \bibinfo{author}{G.~A.~H. Schober},
  \bibinfo{title}{{\itshape Response Theory of the electron-phonon coupling}},
  \bibinfo{howpublished}{arXiv:1606.00012 [cond-mat.mtrl-sci]},
  \bibinfo{year}{2016}{\natexlab{b}}.
\bibitem[{Starke and Schober(2017{\natexlab{a}})}]{EDLor}
\bibinfo{author}{R.~Starke}, \bibinfo{author}{G.~A.~H. Schober},
\newblock \bibinfo{title}{{\itshape Covariant response theory and the boost
  transform of the dielectric tensor}},
\newblock \bibinfo{journal}{Int. J. Mod. Phys. D} \bibinfo{volume}{{\bfseries
  26}} (\bibinfo{year}{2017}{\natexlab{a}}) \bibinfo{pages}{1750163}.
  \bibinfo{note}{{See also arXiv:1702.06985 [physics.class-ph]}}.
\bibitem[{Starke and Schober(2017{\natexlab{b}})}]{EDWave}
\bibinfo{author}{R.~Starke}, \bibinfo{author}{G.~A.~H. Schober},
\newblock \bibinfo{title}{{\itshape Linear electromagnetic wave equations in
  materials}},
\newblock \bibinfo{journal}{Phot. Nano. Fund. Appl.}
  \bibinfo{volume}{{\bfseries 26}} (\bibinfo{year}{2017}{\natexlab{b}})
  \bibinfo{pages}{41}. \bibinfo{note}{{See also arXiv:1704.06615
  [cond-mat.mtrl-sci]}}.
\bibitem[{Schober and Starke(2017)}]{EDFullGF}
\bibinfo{author}{G.~A.~H. Schober}, \bibinfo{author}{R.~Starke},
  \bibinfo{title}{{\itshape General form of the full electromagnetic Green
  function in materials physics}}, \bibinfo{howpublished}{arXiv:1704.07594
  [physics.class-ph]}, \bibinfo{year}{2017}.
\bibitem[{Starke and Schober(2018)}]{EDFresnel}
\bibinfo{author}{R.~Starke}, \bibinfo{author}{G.~A.~H. Schober},
\newblock \bibinfo{title}{{\itshape Why history matters: ab initio rederivation
  of Fresnel equations confirms microscopic theory of refractive index}},
\newblock \bibinfo{journal}{Optik} \bibinfo{volume}{{\bfseries 157}}
  (\bibinfo{year}{2018}) \bibinfo{pages}{275}. \bibinfo{note}{{See also
  arXiv:1705.11004 [physics.optics]}}.
\bibitem[{Bruus and Flensberg(2004)}]{Bruus}
\bibinfo{author}{H.~Bruus}, \bibinfo{author}{K.~Flensberg},
  \bibinfo{title}{{\itshape Many-body quantum theory in condensed matter
  physics: an introduction}}, \bibinfo{publisher}{Oxford University Press},
  \bibinfo{address}{Oxford}, \bibinfo{year}{2004}.
\bibitem[{Keldysh et~al.(1989)Keldysh, Kirzhnitz, and
  Maradudin}]{KeldyshKirzhnitz}
\bibinfo{author}{L.~V. Keldysh}, \bibinfo{author}{D.~A. Kirzhnitz},
  \bibinfo{author}{A.~A. Maradudin}, \bibinfo{title}{{\itshape The dielectric
  function of condensed systems}}, \textnormal{Modern Problems in Condensed
  Matter Sciences}, \bibinfo{publisher}{Elsevier Science Publishers B.V.},
  \bibinfo{address}{Amsterdam}, \bibinfo{year}{1989}.
\bibitem[{Dolgov and Maksimov(1989)}]{Dolgov}
\bibinfo{author}{O.~V. Dolgov}, \bibinfo{author}{E.~G. Maksimov},
\newblock \bibinfo{title}{{\itshape Chapter 4: The dielectric function of
  crystalline systems}},
\newblock in: \bibinfo{editor}{L.~V. Keldysh}, \bibinfo{editor}{D.~A.
  Kirzhnitz}, \bibinfo{editor}{A.~A. Maradudin} (Eds.),
  \bibinfo{booktitle}{{\itshape The dielectric function of condensed systems}},
  \textnormal{Modern Problems in Condensed Matter Sciences},
  \bibinfo{publisher}{Elsevier Science Publishers B.\,V.},
  \bibinfo{address}{Amsterdam}, \bibinfo{year}{1989}.
\bibitem[{Kantorovich(2004)}]{Kantorovich}
\bibinfo{author}{L.~Kantorovich}, \bibinfo{title}{{\itshape Quantum theory of
  the solid state: an introduction}}, \textnormal{Fundamental Theories of
  Physics}, \bibinfo{publisher}{Springer Science+Business Media},
  \bibinfo{address}{Dordrecht}, \bibinfo{year}{2004}.
\bibitem[{Martin(2008)}]{Martin}
\bibinfo{author}{R.~M. Martin}, \bibinfo{title}{{\itshape Electronic structure:
  basic theory and practical methods}}, \bibinfo{publisher}{Cambridge
  University Press}, \bibinfo{address}{Cambridge}, \bibinfo{year}{2008}.
\bibitem[{Martin and Rothen(2002)}]{MartinRothen}
\bibinfo{author}{P.~A. Martin}, \bibinfo{author}{F.~Rothen},
  \bibinfo{title}{{\itshape Many-body problems and quantum field theory: an
  introduction}}, \bibinfo{publisher}{Springer-Verlag},
  \bibinfo{address}{Berlin/Heidelberg}, \bibinfo{year}{2002}.
\bibitem[{Vorwerk et~al.(2016)Vorwerk, Cocchi, and Draxl}]{Draxl16}
\bibinfo{author}{C.~Vorwerk}, \bibinfo{author}{C.~Cocchi},
  \bibinfo{author}{C.~Draxl},
\newblock \bibinfo{title}{{\itshape LayerOptics: microscopic modeling of
  optical coefficients in layered materials}},
\newblock \bibinfo{journal}{Comput. Phys. Commun.} \bibinfo{volume}{{\bfseries
  201}} (\bibinfo{year}{2016}) \bibinfo{pages}{119}.
\bibitem[{R\"omer(2005)}]{Roemer}
\bibinfo{author}{H.~R\"omer}, \bibinfo{title}{{\itshape Theoretical optics: an
  introduction}}, \bibinfo{publisher}{Wiley-VCH Verlag GmbH {\&} Co. KGaA},
  \bibinfo{address}{Weinheim}, \bibinfo{year}{2005}.
\bibitem[{Agranovich and Ginzburg(1984)}]{Agranovich}
\bibinfo{author}{V.~M. Agranovich}, \bibinfo{author}{V.~L. Ginzburg},
\newblock \bibinfo{title}{{\itshape Crystal optics with spatial dispersion, and
  excitons}},
\newblock volume~\bibinfo{volume}{42} of
  \textit{\bibinfo{series}{{\textnormal{Springer Series in Solid-State
  Sciences}}}}, \bibinfo{edition}{2nd} ed.,
  \bibinfo{publisher}{Springer-Verlag}, \bibinfo{address}{Berlin/Heidelberg},
  \bibinfo{year}{1984}.
\bibitem[{Platzmann and Wolff(1973)}]{Platzmann}
\bibinfo{author}{P.~M. Platzmann}, \bibinfo{author}{P.~A. Wolff},
  \bibinfo{title}{{\itshape Waves and interactions in solid state plasmas}},
  \bibinfo{publisher}{Academic Press}, \bibinfo{address}{New York/London},
  \bibinfo{year}{1973}.
\bibitem[{Melrose and McPhedran(1991)}]{Melrose}
\bibinfo{author}{D.~B. Melrose}, \bibinfo{author}{R.~C. McPhedran},
  \bibinfo{title}{{\itshape Electromagnetic processes in dispersive media: a
  treatment based on the dielectric tensor}}, \bibinfo{publisher}{Cambridge
  University Press}, \bibinfo{address}{Cambridge}, \bibinfo{year}{1991}.
\bibitem[{Veselago(1968)}]{Veselago}
\bibinfo{author}{V.~G. Veselago},
\newblock \bibinfo{title}{{\itshape The electrodynamics of substances with
  simultaneously negative values of $\epsilon$ and $\mu$}},
\newblock \bibinfo{journal}{Sov. Phys. Usp.} \bibinfo{volume}{{\bfseries 10}}
  (\bibinfo{year}{1968}) \bibinfo{pages}{509}.
\bibitem[{Born and Wolf(1999)}]{BornWolf}
\bibinfo{author}{M.~Born}, \bibinfo{author}{E.~Wolf}, \bibinfo{title}{{\itshape
  Principles of optics: electromagnetic theory of propagation, interference and
  diffraction of light}}, \bibinfo{edition}{7th} ed.,
  \bibinfo{publisher}{Cambridge University Press},
  \bibinfo{address}{Cambridge}, \bibinfo{year}{1999}.
\bibitem[{Lipson et~al.(2011)Lipson, Lipson, and Lipson}]{Lipson}
\bibinfo{author}{A.~Lipson}, \bibinfo{author}{S.~G. Lipson},
  \bibinfo{author}{H.~Lipson}, \bibinfo{title}{{\itshape Optical Physics}},
  \bibinfo{edition}{4th} ed., \bibinfo{publisher}{Cambridge University Press},
  \bibinfo{address}{Cambridge}, \bibinfo{year}{2011}.
\bibitem[{Landau and Lifshitz(1984)}]{Landau}
\bibinfo{author}{L.~D. Landau}, \bibinfo{author}{E.~M. Lifshitz},
  \bibinfo{title}{{\itshape Electrodynamics of continuous media}},
  volume~\bibinfo{volume}{8} of \textit{\bibinfo{series}{\textnormal{Course of
  Theoretical Physics}}}, \bibinfo{edition}{2nd} ed.,
  \bibinfo{publisher}{Pergamon Press Ltd.}, \bibinfo{address}{Oxford},
  \bibinfo{year}{1984}.
\bibitem[{Zvezdin and Kotov(1997)}]{Zvezdin}
\bibinfo{author}{A.~K. Zvezdin}, \bibinfo{author}{V.~A. Kotov},
  \bibinfo{title}{{\itshape Modern magnetooptics and magnetooptical
  materials}}, \textnormal{Studies in Condensed Matter Physics},
  \bibinfo{publisher}{IOP Publishing Ltd.}, \bibinfo{address}{Bristol},
  \bibinfo{year}{1997}.
\bibitem[{Bredov et~al.(1985)Bredov, Rumyantsev, and Toptygin}]{Bredov}
\bibinfo{author}{M.~M. Bredov}, \bibinfo{author}{V.~V. Rumyantsev},
  \bibinfo{author}{I.~N. Toptygin}, \bibinfo{title}{{\itshape Klassicheskaya
  elektrodinamika}}, \bibinfo{publisher}{Nauka}, \bibinfo{address}{Moscow},
  \bibinfo{year}{1985}.
\bibitem[{Bergmann and Schaefer(1999)}]{Bergmann}
\bibinfo{author}{L.~Bergmann}, \bibinfo{author}{C.~Schaefer},
  \bibinfo{title}{{\itshape Optics of waves and particles}},
  \bibinfo{publisher}{Walter de Gruyter}, \bibinfo{address}{Berlin},
  \bibinfo{year}{1999}.
\bibitem[{Kresse and Furthm\"uller(1996)}]{VASP}
\bibinfo{author}{G.~Kresse}, \bibinfo{author}{J.~Furthm\"uller},
\newblock \bibinfo{title}{{\itshape Efficient iterative schemes for ab initio
  total-energy calculations using a plane-wave basis set}},
\newblock \bibinfo{journal}{Phys. Rev. B} \bibinfo{volume}{{\bfseries 54}}
  (\bibinfo{year}{1996}) \bibinfo{pages}{11169}.
\bibitem[{Blaha et~al.(2001)Blaha, Schwarz, Madsen, Kvasnicka, and
  Luitz}]{Wien2k}
\bibinfo{author}{P.~Blaha}, \bibinfo{author}{K.~Schwarz},
  \bibinfo{author}{G.~Madsen}, \bibinfo{author}{D.~Kvasnicka},
  \bibinfo{author}{J.~Luitz}, \bibinfo{title}{{\itshape WIEN2k, an augmented
  plane wave + local orbitals program for calculating crystal properties}},
  \bibinfo{publisher}{Karlheinz Schwarz}, \bibinfo{address}{Techn.
  Universit\"at Wien, Austria}, \bibinfo{year}{2001}.
\bibitem[{Giannozzi et~al.(2009)Giannozzi, Baroni, Bonini, Calandra, Car,
  Cavazzoni, Ceresoli, Chiarotti, Cococcioni, Dabo et~al.}]{QE-2009}
\bibinfo{author}{P.~Giannozzi}, \bibinfo{author}{S.~Baroni},
  \bibinfo{author}{N.~Bonini}, \bibinfo{author}{M.~Calandra},
  \bibinfo{author}{R.~Car}, \bibinfo{author}{C.~Cavazzoni},
  \bibinfo{author}{D.~Ceresoli}, \bibinfo{author}{G.~L. Chiarotti},
  \bibinfo{author}{M.~Cococcioni}, \bibinfo{author}{I.~Dabo}, et~al.,
\newblock \bibinfo{title}{{\itshape QUANTUM ESPRESSO: a modular and open-source
  software project for quantum simulations of materials}},
\newblock \bibinfo{journal}{J. Phys. Condens. Matter}
  \bibinfo{volume}{{\bfseries 21}} (\bibinfo{year}{2009})
  \bibinfo{pages}{395502}.
\bibitem[{Schwalbe et~al.(2016)Schwalbe, Wirnata, Starke, Schober, and
  Kortus}]{Schwalbe}
\bibinfo{author}{S.~Schwalbe}, \bibinfo{author}{R.~Wirnata},
  \bibinfo{author}{R.~Starke}, \bibinfo{author}{G.~A.~H. Schober},
  \bibinfo{author}{J.~Kortus},
\newblock \bibinfo{title}{{\itshape Ab initio electronic structure and optical
  conductivity of bismuth tellurohalides}},
\newblock \bibinfo{journal}{Phys. Rev. B} \bibinfo{volume}{{\bfseries 94}}
  (\bibinfo{year}{2016}) \bibinfo{pages}{205130}.

\end{thebibliography}

\end{document}